\documentclass[authoryear,3p,preprint]{elsarticle}

\usepackage{amssymb}
\usepackage{amsmath}
\usepackage{booktabs}
\usepackage{xcolor}
\usepackage{float}
\usepackage{subfig}
\usepackage{graphics}
\usepackage{enumitem}
 \usepackage{amsthm}
\usepackage{listings}
\theoremstyle{definition}

\usepackage{natbib}
\usepackage{hyperref}
\hypersetup{
    colorlinks
}

\journal{Computational Statistics \& Data Analysis}

\date{}

\begin{document}

\begin{frontmatter}



\title{Locally  weighted  minimum  contrast estimation for spatio-temporal log-Gaussian Cox processes}

 \author[label1]{Nicoletta D'Angelo\corref{cor1}}
 \author[label1]{Giada Adelfio}
  \author[label2]{Jorge Mateu}
 \address[label1]{Department of Economics, Business, and Statistics, University of Palermo, Palermo, Italy}
\address[label2]{Department of Mathematics, University Jaume I, Castellon, Spain}
\cortext[cor1]{Corresponding author: \texttt{nicoletta.dangelo@unipa.it}}

\begin{abstract}
We propose a local version of spatio-temporal log-Gaussian Cox processes using Local Indicators of Spatio-Temporal Association (LISTA) functions into the minimum contrast procedure to obtain space as well as time-varying parameters. 
 We resort to the joint minimum contrast fitting method to estimate the set of second-order parameters. This approach has the advantage of being suitable in both separable and non-separable parametric specifications of the correlation function of the underlying Gaussian Random Field. 
 We present simulation studies to assess the performance of the proposed fitting procedure, and show an application to seismic spatio-temporal point pattern data.
\end{abstract}

\begin{keyword}
 Local models \sep log-Gaussian Cox processes \sep Minimum contrast
\sep Second-order characteristics \sep  Spatio-temporal point processes  


\end{keyword}

\end{frontmatter}



\section{Introduction}

\label{sec:intro}

Interest in methods for analysing spatial and spatio-temporal point processes is increasing across many fields of science, notably in ecology, epidemiology, geoscience, astronomy, econometrics, and crime research \citep{baddeley:rubak:tuner:15,diggle:13}. When the structure of a given point pattern is observed, it is assumed to come from a realisation of an underlying generating process, whose properties are estimated and then used to describe the structure of the observed pattern. 
The first step in analysing a point pattern is to learn about its first-order characteristics, studying the relationship of the points with the underlying environmental variables that  describe the observed heterogeneity. 
When the purpose of the analysis is to describe the possible interaction among points, that is, if  the given data exhibit spatial inhibition or aggregation, the second-order properties of the process are analysed.
However, in the analysis of spatial (and spatio-temporal) point process data it can be difficult to disentangle the two previous aspects, i.e. the heterogeneity corresponding to spatial variation of the intensity and the dependence structure amongst the points \citep{illian:penttinen:stoyan:stoyan:08,diggle:13}. For this reason, it is attractive and motivating to define and estimate models that account simultaneously for the dependence structure among events, including also the effect of the observed covariates.

Cox processes are a class of models for point phenomena that are environmentally driven and have a clustered structure. They are Poisson processes with a random intensity function depending on unobservable external factors. Two notable classes of Cox point processes are the shot-noise Cox processes \citep{moller:03} and the log-Gaussian Cox processes (LGCPs) \citep{moller1998log}.

LGCPs are arguably the most prominent clustering models, for their flexibility and relatively tractability for describing spatial and spatio-temporal correlated phenomena specifying the moments of an underlying Gaussian Random Field (GRF). Therefore, the main interest is in the estimation of the first and second-order characteristics of the process that depend on the moments of the GRF.

In both purely spatial and spatio-temporal studies, the choice of the estimation procedure depends on several aspects that relate to the application context, the goals of the analysis, and the required computational time. \cite{diggle2013spatial} discusses the three main estimation methods (moment-based, maximum likelihood and Bayesian estimation) for LGCPs.
\cite{siino2018joint}  describe earthquake sequences comparing several Cox model specifications (with separable and non-separable spatio-temporal covariance functions) estimating parameters through the minimum contrast method. 

The aforementioned models can be referred to as `global' models, as they are globally defined and the process properties and estimated parameters are assumed to be constant all through the study area. However, a model with constant parameters, may not adequately represent detailed local variations in the data, since the pattern may present spatial and temporal variations due to the influence of covariates, the scale or spacing between points, and also perhaps due to the abundance of points.
Indeed,  a different way of analysing a point pattern can be based on local techniques  identifying  specific and undiscovered local structure, for instance sub-regions characterised by  different interactions among points, intensity and influence of covariates. 

On one hand, local second-order statistics have been used to obtain further insight into the local structure of the analysed point pattern.
While the use of global spatio-temporal second-order summary statistics is a well-established practice to describe global interaction structures between points in a point pattern \citep{ripley:77,illian:penttinen:stoyan:stoyan:08,gabriel:diggle:09}, the use of local tools was firstly advocated by \cite{siino2018testing}, introducing the  Local Indicators of Spatio-Temporal Association (LISTA) functions as an extension of the purely spatial Local Indicators, whose definition was  given by \cite{anselin:95}.
Successively, \cite{adelfio2020some} introduced local versions of both the homogeneous and inhomogeneous spatio-temporal $K$-functions, and used them as diagnostic tools, while also retaining for local information, showing that the local inhomogeneous $K$-functions can be helpful  to assess the goodness-of-fit of different spatio-temporal models, with the advantage of not relying on any
particular model assumption on the data.

On the other hand, the literature about local models for point processes is quite rare.
For spatial point processes, \cite{baddeley:2017local} presents a general framework based on the local composite likelihood to detect and model gradual spatial variation in any parameter of a spatial stochastic model (such as Poisson, Gibbs and Cox processes). In particular, the parameters in the model that govern the intensity, the dependence of the intensity on the covariates and the spatial interaction between points, are estimated locally. Moreover, this approach has the advantage to detect and model spatial variation in any property of a point process, within a formal likelihood framework providing space-varying parameter estimates, confidence intervals and hypothesis tests. 

\cite{dangelo2021locall} showed that \citeauthor{baddeley:2017local}'s purely local models provide good inferential results by applying them to earthquake data.
However, that work did not account for the temporal dimension of the seismic events, whose realisations depend on their past history, as proved by the existence of  aftershocks. 
Motivated by this,  we propose a local version of spatio-temporal LGCPs employing LISTA functions plugged into the minimum contrast procedure to obtain space as well as time-varying parameters. 
For the parameters of the deterministic part, we follow \cite{baddely:turner:00}, using a quadrature scheme for the locally weighted log-linear Poisson regression.
For the parameters of the coviariance structure, we resort to the joint minimum contrast fitting method proposed by \cite{siino2018joint} to estimate the set of second-order parameters of the spatio-temporal LGCPs. This approach has the advantage of being suitable in both separable and non-separable parametric specifications of the correlation function of the underlying GRF. 

We therefore introduce the local estimation, obtaining a whole set of parameters for each point of the analysed dataset, and we refer to this new procedure as \textit{locally weighted spatio-temporal minimum contrast}.
We show some simulations, finding promising results as the estimates tend to be quite precise on average if compared to the 'global' counterpart, while also reflecting the assumed variability in space and time. 
To enforce these results, we apply the proposed methodology to a real dataset in seismicity, considering the quite convenient separable covariance function of the LGCP.
All the codes are written in the software \cite{R} language, and are available from the first author.
  
The structure of the paper is as follows. Section \ref{sec:stpp} is devoted to the introduction of basic definitions of spatio-temporal point processes. Section \ref{sec:lgcps} briefly reviews the class of spatio-temporal LGCPs. In Section \ref{sec:proposal}, the proposed local minimum contrast estimation based on the LISTA functions is presented.
Section \ref{sec:diagnostics} reports the diagnostic procedure used to assess the goodness-of-fit of a global LGCP, and a proposed modification to deal with local LGCPs.
Then, the performance of the proposed local estimation method is assessed in Section \ref{sec:simulations} through a simulation study.
An application to real seismic data comes in Section \ref{sec:application}. Finally, the paper ends with some conclusions in Section \ref{sec:conclusions}.

\section{Spatio-temporal point processes and main characteristics}
\label{sec:stpp}

We consider a spatio-temporal point process with no multiple points as a random countable subset $X$ of $\mathbb{R}^2 \times \mathbb{R}$, where a point $(\textbf{u}, t) \in X$ corresponds to an event at $ \textbf{u} \in \mathbb{R}^2$ occurring at time $t \in \mathbb{R}$.
A typical realisation of a spatio-temporal point process $X$ on $\mathbb{R}^2 \times \mathbb{R}$ is a finite set $\{(\textbf{u}_i, t_i)\}^n_{
i=1}$ of distinct points within a
bounded spatio-temporal region $W \times T \subset \mathbb{R}^2 \times \mathbb{R}$, with area $|W| > 0$ and length $|T| > 0$, where $n \geq 0$ is not fixed in
advance. In this context, $N(A \times B)$ denotes the number of points of a set $(A \times B) \cap X$, where $A \subseteq W$ and $B \subseteq T$. As usual \citep{daley:vere-jones:08}, when $N(W \times T) < \infty $ with probability 1, which holds e.g. if $X$ is defined on a bounded set, we call $X$ a finite spatio-temporal point process.

For a given event $(\textbf{u}, t)$, the events that are close to $(\textbf{u}, t)$ in both space and time, for each spatial distance $r$ and time lag $h$, are given by the corresponding spatio-temporal cylindrical neighbourhood of the event $(\textbf{u}, t)$, which can be expressed by the Cartesian product as
$$
b((\textbf{u}, t), r, h) = \{(\textbf{v}, s) : ||\textbf{u} - \textbf{v}|| \leq r, |t - s| \leq h\} , \quad \quad
(\textbf{u}, t), (\textbf{v}, s) \in W \times T,
$$
where $||\cdot||$ denotes the Euclidean distance in $\mathbb{R}^2$. Note that $b((\textbf{u}, t), r, h)$ is a cylinder with centre (\textbf{u}, t), radius $r$, and height $2h$.

Product densities $\lambda^{(k)}, k  \in \mathbb{N} \text{ and }  k  \geq 1 $, arguably the main tools in the statistical analysis of point processes, may be defined through the so-called Campbell Theorem (see \cite{daley:vere-jones:08}), which states that given a spatio-temporal point process $X$, and for any non-negative function $f$ on $( \mathbb{R}^2 \times \mathbb{R} )^k$, we have
\begin{equation*}
  \mathbb{E} \Bigg[ \sum_{\zeta_1,\dots,\zeta_k \in X}^{\ne} f( \zeta_1,\dots,\zeta_k)\Bigg]=\int_{\mathbb{R}^2 \times \mathbb{R}} \dots \int_{\mathbb{R}^2 \times \mathbb{R}} f(\zeta_1,\dots,\zeta_k) \lambda^{(k)} (\zeta_1,\dots,\zeta_k) \prod_{i=1}^{k}\text{d}\zeta_i,
\label{eq:campbell0}  
\end{equation*}
that constitutes an essential result in spatio-temporal point process theory. In particular, for $k=1$ and $k=2$, these functions are respectively called the \textit{(first-order) intensity function} $\lambda$ and the \textit{(second-order) product density} $\lambda^{(2)}$.
Broadly speaking, the intensity function describes the rate at which the events occur in the given spatio-temporal region, while the second-order product densities are used when the interest is in describing spatio-temporal variability and correlations between pair of points of a pattern. They represent the point process analogues of the mean function and the covariance function of a real-valued process, respectively.
Then, the intensity function is defined as 
\begin{equation*}
 \lambda(\textbf{u},t)=\lim_{|\text{d}\textbf{u} \times \text{d}t| \rightarrow 0} \frac{\mathbb{E}[N(\text{d}\textbf{u} \times \text{d}t )]}{|\text{d}\textbf{u} \times \text{d}t|},  
\end{equation*}
where $\text{d}\textbf{u} \times \text{d}t $ defines a small region around the point $(\textbf{u},t)$ and $|\text{d}\textbf{u} \times \text{d}t|$ is its volume. The second-order intensity function is
\begin{equation*}
     \lambda^{(2)}((\textbf{u},t),(\textbf{v},s))=\lim_{|\text{d}\textbf{u} \times \text{d}t|,|\text{d}\textbf{v} \times \text{d}s| \rightarrow 0} \frac{\mathbb{E}[N(\text{d}\textbf{u} \times \text{d}t )N(\text{d}\textbf{v} \times \text{d}s )]}{|\text{d}\textbf{u} \times \text{d}t||\text{d}\textbf{v} \times \text{d}s|}.
\end{equation*}
Finally, the pair correlation function
\begin{equation*}
    g((\textbf{u},t),(\textbf{v},s))=\frac{ \lambda^{(2)}((\textbf{u},t),(\textbf{v},s))}{\lambda(\textbf{u},t)\lambda(\textbf{v},s)}
\end{equation*}
can be interpreted formally as the standardised probability density that an event occurs in each of two small volumes, $\text{d}\textbf{u} \times \text{d}t$ and $\text{d}\textbf{v} \times \text{d}s$, in the sense that for a Poisson process, $g((\textbf{u},t),(\textbf{v},s))=1$.

\section{Spatio-temporal log-Gaussian Cox processes}
\label{sec:lgcps}

	In the Euclidean context, LGCPs are one of the most prominent clustering models. By specifying the intensity of the process and the moments of the underlying GRF, it is possible to estimate both the first and second-order characteristics of the process. 
	Following the inhomogeneous specification in \cite{diggle2013spatial}, a  LGCP for a generic point in space and time has the intensity
	\begin{equation*}
		\Lambda(\textbf{u},t)=\lambda(\textbf{u},t)\exp(S(\textbf{u},t))
	\end{equation*}
where $S$ is a Gaussian process with $\mathbb{E}(S(\textbf{u},t))=\mu=-0.5\sigma^2$ and so  $\mathbb{E}(\exp{S(\textbf{u},t)})=1$ and with variance and covariance matrix $\mathbb{C}(S(\textbf{u}_i,t_i),S(\textbf{u}_j,t_j))=\sigma^2 \gamma(r,h)$ under the stationary assumption, with $\gamma(\cdot)$ the correlation function of the GRF, and $r$ and $h$ some spatial and temporal distances. Following \cite{moller1998log}, the first-order product density and the pair correlation function of an LGCP are $\mathbb{E}(\Lambda(\textbf{u},t))=\lambda(\textbf{u},t)$ and $g(r,h)=\exp(\sigma^2\gamma(r,h))$, respectively.  
		In this paper, we first consider a separable structure for the covariance function of the GRF \citep{brix2001spatiotemporal} that has exponential form for both the spatial and the temporal components,
	\begin{equation}
		\mathbb{C}(r,h)=\sigma^2\exp \bigg(\frac{-r}{\alpha}\bigg)\exp\bigg(\frac{-h}{\beta}\bigg),
		\label{eq:cov}
	\end{equation}
where $\sigma^2$ is the variance, $\alpha$ is the scale parameter for the spatial distance and $\beta$ is the scale parameter for the temporal one.
The exponential form is widely used in this context and nicely reflects the decaying correlation structure with distance or time.
Moreover, we may consider a non-separable covariance of the GRF useful to describe
more general situations.
\cite{gneiting2006geostatistical} review parametric non-separable space-time covariance functions for geostatistical models. Following the parametrisation in \cite{schlather2015analysis}, Gneiting covariance function can be written as
$$
		\mathbb{C}(r,h) = (\psi(h) + 1)^{ - d/2} \varphi \bigg( \frac{r}{\sqrt{\psi(h) + 1}}  \bigg) \qquad r \geq 0,  \quad h \geq 0, 
$$
where $\varphi(\cdot)$ is a complete monotone function associated to
the spatial structure, and $\psi(\cdot)$ is a positive function with a
completely monotone derivative associated to the temporal
structure of the data. For example, the choice $d = 2$,
$\varphi(r)=\sigma^2 \exp ( - (\frac{r}{\alpha})^{\gamma_s})$ and 
$\psi(h)=((\frac{h}{\beta})^{\gamma_t} + 1)^{\delta/\gamma_t}$
yields to the parametric family
	\begin{equation}
		\mathbb{C}(r,h) = \frac{\sigma^2}{((\frac{h}{\beta})^{\gamma_t} + 1)^{\delta/\gamma_t}} \exp \Biggl( - \frac{(\frac{r}{\alpha})^{\gamma_s}}{((\frac{h}{\beta})^{\gamma_t} + 1)^{\delta/(2\gamma_t)}} \Biggl),
		\label{eq:nonsep}
	\end{equation}
where $\alpha > 0$ and $\beta > 0$ are scale parameters of space and time, $\delta$ takes values in $(0, 2]$, and $\sigma^2$ is the variance.
Another parametric covariance belongs to the Iaco-Cesare family \citep{de2002fortran,de2002nonseparable}, and there is a wealth of covariance families that could well be used for our purposes.

\section{Model estimation}
\label{sec:proposal}

Driven by a GRF, controlled in turn by a specified covariance structure, the implementation of the LGCP framework in practice requires a proper estimate of the intensity function.
In general, the Cox model is estimated by a two-step procedure, involving first the intensity and then the cluster or correlation parameters. First, a Poisson process with a particular model for the log-intensity is fitted to the point pattern data, providing the estimates of the coefficients of all the terms that characterise the intensity. 
Then, the estimated  intensity is  taken as the true one and the cluster  or correlation parameters are estimated using either the \textit{method of minimum contrast} \citep{pfanzagl1969measurability,eguchi1983second,diggle1979parameter,diggle1984monte,moller1998log,davies2013assessing,siino2018joint}, \textit{Palm likelihood} \citep{ogata1991maximum,tanaka2008parameter}, or
 \textit{composite likelihood} \citep{guan2006composite}.
   	The most  common technique is the \textit{minimum contrast}, and it is the method which we shall refer to here.
   
In the following, we first review and extend the  statistical theory and computational strategies for local estimation and inference for spatio-temporal Poisson point processes, in Section \ref{sec:quad}.
Then, in Section \ref{sec:method1}  we recall the joint minimum contrast procedure, that we extend in Section \ref{sec:proposall} to the local context.

\subsection{Estimating the first-order intensity function through local likelihood}
\label{sec:quad}

We assume that the template model is a Poisson process, with a parametric intensity or rate function $$\lambda(\textbf{u}, t; \theta), \quad  X \in
W,\quad  t \in T, \quad \theta \in \Theta.$$ The log-likelihood is 
\begin{equation}
    \log L(\theta) = \sum_i
\lambda(\textbf{u}_i, t_i; \theta) - \int_W\int_T
\lambda(\textbf{u}, t; \theta) \text{d}t\text{d}u
\label{eq:glo_lik}
\end{equation}
up to an additive constant, where the sum is over all points $\textbf{u}_i$ in $X$. 
We often consider intensity models of log-linear form
\begin{equation}
   \lambda(\textbf{u}, t; \theta) = \exp(\theta Z(\textbf{u}, t) + B(\textbf{u},t )), \quad
\textbf{u} \in W,\quad  t \in T
\label{eq:glo_mod}
\end{equation}
where $Z(\textbf{u}, t)$ is a vector-valued covariate function, and $B(\textbf{u}, t)$ is a scalar offset. 

\subsubsection{Quadrature scheme}

Following \cite{berman1992approximating}, we use a finite quadrature approximation to the log-likelihood, as implemented in the \texttt{spatstat} package \citep{spat}. Renaming the data points as $\textbf{x}_1,\dots , \textbf{x}_n$ with $(\textbf{u}_i,t_i) = \textbf{x}_i$ for $i = 1, \dots , n$, then generate $m$  additional ‘‘dummy points’’ $(\textbf{u}_{n+1},t_{n+1}) \dots , (\textbf{u}_{m+n},t_{m+n})$ to
form a set of $n + m$ quadrature points (where $m > n$). Then we determine quadrature weights $a_1, \dots , a_m$
so that integrals in \eqref{eq:glo_lik} can be approximated by a Riemann sum
\begin{equation}
    \int_W \int_T \lambda(\textbf{u},t;\theta)\text{d}t\text{d}u \approx \sum_{k = 1}^{n + m}a_k\lambda(\textbf{u},t;\theta)
    \label{eq:riemann}
\end{equation}
where $a_k$ are the quadrature weights such that $\sum_{k = 1}^{n + m}a_k = l(W \times T)$ where $l$ is the Lebesgue measure.

Then the log-likelihood \eqref{eq:glo_lik} of the template model can be approximated by
\begin{equation*}
        \log L(\theta)   \approx
\sum_i
\log \lambda(\textbf{x}_i; \theta) +
\sum_j
(1 - \lambda(\textbf{u}_j,t_j; \theta))a_j
 =
\sum_j
e_j \log \lambda(\textbf{u}_j, t_j; \theta) + (1 - \lambda(\textbf{u}_j, t_j; \theta))a_j
\end{equation*}

where $e_j = 1\{j \leq n\}$ is the indicator that equals $1$ if $u_j$ is a data point. Writing $y_j = e_j/a_j$ this becomes
\begin{equation}
    \log L(\theta) \approx
\sum_j
a_j
(y_j \log \lambda(\textbf{u}_j, t_j; \theta) - \lambda(\textbf{u}_j, t_j; \theta))
+
\sum_j
a_j.
\label{eq:approx}
\end{equation}

Apart from the constant $\sum_j a_j$, this expression is formally equivalent to the weighted log-likelihood of
a Poisson regression model with responses $y_j$ and means $\lambda(\textbf{u}_j,t_j; \theta) = \exp(\theta Z(\textbf{u}_j,t_j) + B(\textbf{u}_j,t_j))$. This can be
maximised using standard GLM software. For more details see \cite{berman1992approximating,baddeley2000non,baddely:turner:moller:05} Section 9.8.

We define the spatio-temporal quadrature scheme by defying a spatio-temporal partition of $W \times T$ into cubes $C_k$ of equal volume $\nu$, assigning the weight $a_k=\nu/n_k$  to each quadrature point (dummy or data) where $n_k$ is the number of points that lie in the same cube as the point $u_k$ \citep{raeisi2021spatio}. 

The number of dummy points should be sufficient for an accurate estimate of the likelihood. Following \cite{baddeley2000non} and \cite{raeisi2021spatio}, we start with a number of dummy points $m \approx 4 n$, increasing it until $\sum_k a_k = l(W \times T)$.

\subsubsection{Local Poisson models in space and time}
\label{sec:weight}
The local log-likelihood associated with the  spatio-temporal location $(\textbf{v},s)$ is given by \begin{equation}
      \log L((\textbf{v},s);\theta) = \sum_i w_{\sigma_s}(\textbf{u}_i - \textbf{v}) w_{\sigma_t}(t_i - s)
\lambda(\textbf{u}_i, t_i; \theta)  - \int_W \int_T
\lambda(\textbf{u}, t; \theta) w_{\sigma_s}(\textbf{u}_i - \textbf{v}) w_{\sigma_t}(t_i - s) \text{d}t \text{d}u  
\label{eq:loc_lik}
\end{equation}
where $w_{\sigma_s} $ and $w_{\sigma_t}$ are weight functions, and $\sigma_s, \sigma_t > 0$ are the smoothing bandwidths. It is not
necessary to assume that $w_{\sigma_s}$ and $w_{\sigma_t}$ are probability densities. For simplicity, we shall consider only kernels of fixed
bandwidth, even though spatially adaptive kernels could also be used.
Note that if the template model is the homogeneous Poisson process with intensity $\lambda$, then the local
likelihood estimate $\hat{\lambda}(\textbf{v}, s)$ reduces to the kernel estimator of the point process intensity \citep{diggle:13} with
kernel proportional to $w_{\sigma_s}w_{\sigma_t}$.

We now use in \eqref{eq:loc_lik} a similar approximation as in \eqref{eq:approx} for the local log-likelihood associated
with each desired location $(\textbf{v},s) \in W \times T$
\begin{equation}
 \log L((\textbf{v},s); \theta) \approx
\sum_j
w_j(\textbf{v},s)a_j
(y_j \log \lambda(\textbf{u}_j,t_j; \theta) - \lambda(\textbf{u}_j,t_j; \theta))
+
\sum_j
w_j(\textbf{v},s)a_j   ,
    \label{eq:approx_local}
\end{equation}
where $w_j(\textbf{v},s) = w_{\sigma_s}(\textbf{v} - \textbf{u}_j) w_{\sigma_t}(s - t_j)$.

Basically, for each
desired location $(\textbf{v},s)$, we replace the vector of quadrature weights $a_j$ by $a_j(\textbf{v},s)= w_j(\textbf{v},s)a_j$ where
$w_j (\textbf{v},s) = w_{\sigma_s}(\textbf{v} - \textbf{u}_j)w_{\sigma_t}(s - t_j)$, and use the GLM software to fit the Poisson regression.
The local likelihood is defined at any location $(\textbf{v},s)$ in continuous space. In practice it will be enough to
consider a grid of points $(\textbf{v},s)$. The choice of the grid depends on the computing resources, the computations
required at each location, and the spatial resolution required.

\paragraph{Choice of the bandwidths}

Bandwidth selection is always an unavoidable topic with kernel estimation. In principle, the bandwidth should be selected according to data resolution, which is at the order of 1-10 times of the nearest neighbouring distance \citep{zhuang2020estimation}.
However, the simple kernel estimate with a fixed bandwidth has a serious disadvantage, that is, for a (spatially) clustered point dataset, a small bandwidth gives a noisy estimate for the sparsely populated area, whereas a large bandwidth mixes up the boundaries between the densely populated and sparsely populated areas. Therefore, instead of the kernel estimates $w$ with fixed bandwidth $\sigma$, we can adopt variable kernel estimates where $\sigma_j$ represents the varying bandwidth calculated for each event $j$.
We can determine the variable bandwidth by
$$ \sigma_j = max \{ \epsilon, \inf ( r : N [ B (\textbf{u}_j ; r)] > n_p)  \}, $$
where $\epsilon$ is a small number, $B(\textbf{u}_j;r)$ is the disk centered at $\textbf{u}_j$ with a radius of $r$, and $n_p$ is a positive integer, i.e. $\sigma_j$ is the distance to $n_p$-th closest event.
In other words, once a suitable integer between 10 and 100 for the parameter $n_p$ is chosen, we calculate a bandwidth value $\sigma_j$,
of each event $j$, as the radius of the smallest circle centered at the location of the $j$th event that includes at least $n_p$ other events.
See \cite{silverman:86} for similar locally dependent estimates, \cite{zhuang2002stochastic} for their use in a seismic context, and \cite{musmeci1986variable,choi1999nonparametric} for similar ideas.

\subsection{Estimating the covariance parameters through minimum contrast}
\label{sec:method1}

			The arbitrariness of the minimum contrast procedure can be criticised, however the relative computational simplicity with respect to other estimation procedures (such as likelihood or Bayesian estimation procedures) makes this method suitable for estimating Cox process parameters \citep{siino2018joint}.
	 The procedure selects those parameters that minimise the squared discrepancy between parametric and non-parametric representations of the second-order properties of the LGCP. Minimum contrast estimation is used because direct likelihood-based inference for the parameters of interest is generally not possible.	
	It is important to notice that the intuitiveness of the minimum contrast procedure is offset by the numerous subjective decisions that must be made in order to implement the criterion in practice.
	
	Let the function $J$ represent either the pair correlation function $g$ or the $K$-function, and $\hat{J}$ stands for the corresponding non-parametric estimate. The minimum contrast estimates $\hat{\sigma}^2$ and $\hat{\alpha}$ are found minimising
\begin{equation}
   M_J\{ \sigma^2,\alpha\}=\int_{r_0}^{r_{max}}\phi(r)\{\nu[\hat{J}(r)]-\nu[J(r;\sigma^2,\alpha)]\}^2 \text{d}r
\label{eq:mj}
\end{equation}
	where $r_0$ and $r_{max}$ are the lower and upper lag limits of the contrast criterion, $\phi$ denotes some scalar weight associated with each spatial lag $r$, $\nu$ represents some transformation of its argument and the approximation of $M_J$ is obtained by summing over a fine sequence of lags $R=\{r_0,r_1,\dots,r_{max}\}$ equally spaced, so that $R_{diff}=r_b-r_a, b>a$
$$
M_J\{ \sigma^2,\alpha\} \approx R_{diff}^{-1}\sum_{r\in R}\phi(r)\{\nu[\hat{J}(r)]-\nu[J(r;\sigma^2,\alpha)]\}^2.
$$
	Estimation of $\tau$ is performed minimising the squared discrepancy between the covariance function between the expected frequency of observations at two points in time $C(t,t-h;\tau)$ and its natural estimator, the empirical autocovariance function $\hat{C}(t,t-h)$, for a finite sequence  of temporal lags.
	The contrast  criterion for the temporal part is 
\begin{equation}
	M_C\{\tau\}=\sum_{h=1}^{h_{max}} \sum_{t=h+1}^{T} [\hat{C}(t,t-h)-C(t,t-h;\tau)]^2,
    \label{eq:mc}
\end{equation}
	for some user-specified value of $h_{max}<T$, that it is the temporal counterpart of $r_{max}$. The theoretical version of the temporal
covariance depends upon the spatial parameters; thus the spatial parameters must be estimated first and then plugged into the temporal minimum contrast procedure.

	Alternatively, \cite{siino2018joint} proposed a new fitting method to estimate the set of second-order parameters for the class of LGCPs with constant first-order intensity function.
Hereafter we will denote by $\boldsymbol{\theta}$ the vector of (first-order) intensity parameters, and by $\boldsymbol{\psi}$ the cluster parameters, also denoted as correlation or interaction parameters by some authors. For instance, in the case of a spatio-temporal LGCP with exponential covariance, as the one in Equation \eqref{eq:cov}, the cluster parameters correspond to $\boldsymbol{\psi}= (\sigma, \alpha,  \beta)$.
The second-order parameters  $\boldsymbol{\psi}$  are found by minimising
	\begin{equation}
	M_J\{ \boldsymbol{\psi}\}=\int_{h_0}^{h_{max}} \int_{r_0}^{r_{max}} \phi(r,h) \{\nu[\hat{J}(r,h)]-\nu[J(r,h;\boldsymbol{\psi})]\}^2 \text{d}r \text{d}h,
	\label{eq:mjoint}
	\end{equation}
	where $\phi(r, h)$ is a weight that depends on the space-time distance
and $\nu$ is a transformation function.
With simulations, \cite{siino2018joint} show that the \textit{joint minimum contrast procedure},
based on the spatio-temporal pair correlation function, provides reliable estimates.  Its main advantage is that it can be used in the case of both separable and non-separable
parametric specifications of the correlation function of the underlying GRF. Therefore, it represents a more flexible method with respect to other current available methods, and it is the method that we chose to extend into the local context.

\subsection{Locally weighted spatio-temporal minimum contrast}
\label{sec:proposall}

In the purely spatial context, a localised version of minimum contrast is developed using the local
$K$-functions or local pair correlation functions by \cite{baddeley:2017local},   bearing a very close resemblance to the local Palm
likelihood approach, whose implementation is provided by the function \texttt{locmincon()} of the R package \texttt{spatstat.local} \citep{spat.loc}.

Combining the \textit{joint minimum contrast} \citep{siino2018joint} and the \textit{local minimum contrast} \citep{baddeley:2017local} procedures, we can obtain a vector of parameters $\boldsymbol{\psi}_i$ for each point $i$, by minimising
	\begin{equation}
	M_{J,i}\{ \boldsymbol{\psi}_i \}=\int_{h_0}^{h_{max}}\int_{r_0}^{r_{max}} \phi(r,h) \{ \nu[\bar{J}_i(r,h)]-\nu[J(r,h;\boldsymbol{\psi})]\}^2 \text{d}r \text{d}h,
	\label{eq:mjoint_weight_i}
	\end{equation}
	where $\bar{J}_i(r,h)$ is the average of the local functions $\hat{J}_i(r,h)$, weighted by some point-wise kernel estimates.
This procedure not only provides individual estimates, but it does also account for the vicinity of the observed points, and therefore the contribution of their displacement on the estimation procedure.
 This conceptually resembles the methodology used for the local log-likelihood in Section \eqref{sec:weight}.
Following \cite{siino2018joint}, we suggest using $\phi(r,h)=1$ and  $\nu$ as the identity function. Then, $r_{max}$ and $h_{max}$ are selected as 1/4 of the maximum observable
spatial and temporal distances.

	Thus, consider again the weights $w_i= w_{i,\sigma_s}w_{i,\sigma_t}$ given by some kernel estimates. The same considerations hold for the choice of the bandwidth in the local log-likelihood. Then, the averaged weighted local statistics $\bar{J}_i(r,h)$ in \eqref{eq:mjoint_weight_i}, for each point $i$, is 
	$$
	\bar{J}_i(r,h)= \frac{\sum_{i=1}^{n}\hat{J}_i(r,h)w_i}{\sum_{i=1}^{n}w_i}.
	$$

	In particular, we consider $\hat{J}_i(\cdot)$ as the local spatio-temporal pair correlation function \citep{stpp}, evaluated for each ith event $(\textbf{u}_i,t_i)$,
	\begin{equation}
	   \hat{J}_i(r,h) =\hat{g}_{i}(r,h)= \frac{1}{4 \pi r |W \times T|\hat{\lambda}^2} 
	   \sum_{j \ne i} \frac{\kappa_{\epsilon,\delta}(||\textbf{u}_i-\textbf{u}_j||-r,|t_i-t_j|-h)}{\omega(\textbf{u}_i,\textbf{u}_j)\omega(t_i,t_j)}
	   \label{eq:pcf}
	\end{equation}
where $\omega$ is the edge correction factor.

The kernel function $\kappa$ has a multiplicative form
$\kappa_{\epsilon,\delta}(||\textbf{u}_i-\textbf{u}_j||-r,|t_i-t_j|-h) = \kappa_{\epsilon}(||\textbf{u}_i-\textbf{u}_j||-r)\kappa_{\delta}(|t_i-t_j|-h)$
where $\kappa_{\epsilon}$ and $\kappa_{\delta}$ are kernel functions with
bandwidths $\epsilon$  and $\delta$, respectively. Both of them are computed using the Epanechnikov kernel \citep{illian:penttinen:stoyan:stoyan:08} and the bandwidths are estimated with a
direct plug-in method \citep{sheather:91} using the
\texttt{dpik} function of the package \texttt{Kernsmooth} \citep{KernSmooth}.

\section{Diagnostics}
\label{sec:diagnostics}

For diagnostics of the  proposed LGCP models and the corresponding estimation procedure, a test for spatio-temporal clustering can be employed  as in \cite{tamayo:mateu:diggle:14,siino2018joint,dangelo2021}.
$Q$ realisations from spatio-temporal LGCPs are computed as follows: 	
\begin{enumerate}
\item Generate a realisation from a GRF $S(\textbf{u},t)$, with covariance function \\ $\mathbb{C}((\textbf{u},t),(\textbf{v},s))$ and mean function $\mu(\textbf{u},t)$;
\item Define the generating intensity function  $\lambda_0(\textbf{u},t)=\hat{\lambda}(\textbf{u},t)\exp(S(\textbf{u},t))$;
\item Set an upper bound $\lambda_{max}$ for $\lambda_0(\textbf{u},t)$;
\item Simulate a homogeneous Poisson process $\textbf{x}$ with intensity $\lambda_{max}$ and denote by $N$ the number of generated points, with coordinates $(\textbf{u}', t')$;
\item Compute $p(\textbf{u}',t')=\frac{\lambda(\textbf{u}',t')}{\lambda_{max}}$ for each point $(\textbf{u}',t')$ of a homogeneous Poisson  process;
\item Generate a sample $\textbf{p}$ of size $N$ from the uniform distribution on $(0,1)$; 
\item Thin the simulated homogeneous Poisson process $\textbf{x}$ retaining the $n \leq N$ locations for which $\textbf{p} \leq p(\textbf{u}',t')$. 
\end{enumerate}
The GRF is generated using the function \texttt{RFsim} of the R package \texttt{CompRandFld} \citep{padoan2015analysis}. 
Having simulated $Q>1$ realisations of spatio-temporal point patterns $\textbf{x}_1,\dots,\textbf{x}_Q$, these simulated processes can be used for computing $Q$  inhomogeneous  spatio-temporal $K$-functions as given by \cite{gabriel:diggle:09} 
\begin{equation}
        \hat{K}(r,h)=\frac{|W||T|}{n(n-1)}\sum_{i=1}^n \sum_{j > i} \frac{I(||\textbf{u}_i-\textbf{u}_j||\leq r,|t_i-t_j| \leq h)}{\hat{\lambda}(\textbf{u}_i,t_i)\hat{\lambda}(\textbf{u}_j,t_j)},
    \label{eq:kinh}
\end{equation}
where  $I(\cdot)$ is the indicator function, such that $I(x)=1$ if $x$ is true.
Given the $Q$ inhomogeneous $K$-functions, their corresponding mean and variance, denoted by $E_K$ and $V_K$ respectively, are computed.
The overall test statistic is  
\begin{equation*}
T_q=\int_{r_0}^{r_{max}} \int_{h_0}^{h_{max}} \frac{\hat{K}_q(r,h)-E_K(r,h)}{\sqrt{V_K(r,h)}} ,
\label{eq:T}
\end{equation*}
one for each $\hat{K}_q(r,h)$, obtaining $T_1,\dots,T_{Q}$. $r_{max}$ and $h_{max}$ are the maximum spatial and temporal distances considered for the inhomogeneous $K$-functions. Then, the same test statistic is computed also for the empirical point pattern, and denoted by $T^*$.
The p-value is then defined as
\begin{equation}
\frac{1+\sum_{q=1}^Q I(T_q > T^*)}{Q+1},
\label{eq:pvalue}
\end{equation}	 
basically counting how many times the \textit{K}-functions computed over the simulated processes are higher than the one computed on the original point process. Then, if the obtained p-value is smaller than a significance level $\alpha$ there is  evidence against the null hypothesis, that is, the analysed spatio-temporal point process still presents globally some clustering behaviour.
	
The resulting inhomogeneous $K$-functions can also be used to obtain upper and lower envelopes at a chosen significance level $\alpha$, and so to visually assess the possible residual clustered structure of the analysed point pattern, unexplained by the proposed model. In particular, if the estimated $K$-function lays above  the obtained envelopes, then there is still  a clustering behaviour of points that is not completely described by the proposed model.
Furthermore, by a visual assessment of the results we can also get  indications of  possible ranges needing more complex models, able to explain and take into account the residual spatio-temporal dependence of the data.

To deal with local LGCPs, we provide a slightly modified diagnostics method.
Indeed, given the local covariance parameters, diagnostics can be carried out following the same procedure outlined above, but simulating a \textit{local} GRF from the estimates of the fitted model. This will reflect the variability of the estimated covariance parameters.
From the local GRF we can then generate a local spatio-temporal LGCP. The algorithm is basically the same as the previous one, except that we substitute step 1 by the following additional steps. The rest remains the same.
\begin{itemize}
\item[1.a] Generate a realisation from a GRF $S(\textbf{u},t)$, with covariance function \\ $\mathbb{C}((\textbf{u},t),(\textbf{v},s))$ and mean function $\mu(\textbf{u},t)$;
    \item[1.b] Define a spatio-temporal grid, whose breaks are evenly spaced (the definition of the grid depends on the level of detail to be given), for the simulation of the \textit{local} GRF, and for each point $(\textbf{u}_i,t_i)$, obtain a GRF $S(\textbf{u}_i,t_i)$ simulated from the individual set of estimates $\boldsymbol{\psi}_i$;
    \item[1.c] Compute the average of the GRF values  obtained in 1.a within the sub-grids defined in 1.b;
    \item[1.d] Fill the empty sub-grids with the global GRF values;
\end{itemize}
The procedure 1.a - 1.d ends  obtaining a \textit{local} GRF $S'(\textbf{u},t)$, for which the information of the local estimates is exploited.

\section{Simulation study}
\label{sec:simulations}

A simulation study with a number of scenarios is carried out to assess and compare the performances of the global estimation method \cite{siino2018joint} and the local one proposed in Section \ref{sec:proposall}.

In a first set of scenarios, we assume a stationary and isotropic LGCP with a separable structure of the covariance of the underlying GRF, with an exponential model both in space and time as in Equation \eqref{eq:cov},  the vector of parameters given by $\boldsymbol{\psi} = \{ \sigma^2 ,\alpha   ,\beta \}$.
 For each scenario, $200$ point patterns are generated with $n = 1000$ expected number of points in the spatio-temporal window $W \times T = [0,1]^2 \times [0,50]$, with constant first-order intensity equal to $b = \log (n / |W \times T|)$.
We consider several degrees  of  clustering  in  the  process  with  variance $\sigma^2 = \{ 5, 8\}$ and scale parameters in space and time, $\alpha = \{ 0.005, 0.10, 0.25\}$  and $\beta = \{2,  5, 10\}$. The mean of the GRF is fixed $\mu = -0.5 \sigma^2$. These sets of parameters are the same used in the simulation study in \cite{siino2018joint}. 
Table \ref{tab:sims1} contains  mean and quartiles of the distributions of the estimated local parameters $\hat{\boldsymbol{\psi}_i}= \{ \hat{\sigma}_i^2 ,\hat{\alpha}_i   ,\hat{\beta}_i \}$, averaged over the 200 simulated point patterns.

The results obtained are quite promising: indeed, even considering fixed bandwidths for the weights in the proposed locally weighted minimum contrast, the procedure manage to provide quite precise estimates. This is particularly evident if compared to the results in \cite{siino2018joint} and, even before, in \cite{davies2013assessing}, where the authors provide a number of simulation studied to assess the overall performance of the minimum contrast procedure under different aspects, concluding that the estimates of the variance $\sigma^2$ strongly tend to be underestimated. However, our main goal here is not to provide an alternative to the classical global minimum contrast procedure, but instead to estimate local parameters. This objective is clearly achieved as we manage to obtain a whole distribution for each parameter (of each analysed point pattern).

The same results hold for the scenarios simulated from $200$ LGCPs with $n = 1000$ points each and the non-separable covariance function in \eqref{eq:nonsep}, as evident from Table \ref{tab:sims2}.

\begin{table}[H]
\centering
\caption{\label{tab:sims1} Mean (m) and quartiles of the distributions of the local parameters $\hat{\boldsymbol{\psi}_i}= \{ \hat{\sigma}_i^2 ,\hat{\alpha}_i   ,\hat{\beta}_i \}$, averaged over the 200 simulated point patterns generated assuming an exponential form in both the spatial and temporal dimensions for the GFR with covariance as in \eqref{eq:cov}.}
\resizebox{\textwidth}{!}{\begin{tabular}{ccc|cccc|cccc|cccc}
  \toprule
    & True & &   &  $\hat{\sigma}^2$  &  &  &  & $\hat{\alpha}$ & &     &  & $\hat{\beta}$  &     &  \\
      \midrule
      $\sigma^2$ & $\alpha$ & $\beta$ & 25\% & 50\% & m (mse)  & 75\%  & 25\% & 50\% & m (mse) & 75\%  & 25\% & 50\% & m (mse) & 75\%  \\ 
  \midrule
   5 & 0.05 &   2 & 5.27 & 6.30 & 6.45(2.38) & 7.60 & 0.05 & 0.07 & 0.14(5.26) & 0.09 & 1.77 & 2.26 & 2.63(5.19) & 2.97 \\ 
     & 0.10 &     & 4.64 & 5.51 & 5.67(2.56) & 6.47 & 0.09 & 0.11 & 0.13(0.91) & 0.15 & 1.80 & 2.32 & 2.61(4.80) & 3.08 \\ 
     & 0.25 &     & 3.68 & 4.39 & 4.63(3.11) & 5.37 & 0.19 & 0.24 & 0.34(4.47) & 0.32 & 1.69 & 2.22 & 2.50(5.39) & 2.92 \\ 
     & 0.05 &   5 & 4.36 & 5.43 & 5.54(2.64) & 6.58 & 0.05 & 0.07 & 0.12(4.02) & 0.09 & 3.36 & 4.43 & 5.03(4.74) & 5.93 \\ 
     & 0.10 &     & 4.13 & 4.96 & 5.14(2.95) & 5.97 & 0.09 & 0.11 & 0.14(1.93) & 0.15 & 3.40 & 4.45 & 5.09(4.61) & 6.17 \\ 
     & 0.25 &     & 3.29 & 4.10 & 4.27(4.20) & 5.02 & 0.17 & 0.24 & 0.40(6.56) & 0.34 & 3.04 & 4.21 & 4.89(6.81) & 5.90 \\ 
     & 0.05 &  10 & 4.08 & 5.03 & 5.20(3.09) & 6.12 & 0.05 & 0.06 & 0.10(4.07) & 0.08 & 5.69 & 7.85 & 8.61(7.75) & 10.54 \\ 
     & 0.10 &     & 3.66 & 4.44 & 4.66(4.19) & 5.50 & 0.08 & 0.11 & 0.16(4.23) & 0.14 & 5.64 & 8.00 & 8.85(8.58) & 11.07 \\ 
     & 0.25 &     & 3.05 & 3.73 & 3.97(4.63) & 4.70 & 0.16 & 0.22 & 0.35(6.07) & 0.30 & 5.15 & 7.09 & 8.15(11.02) & 9.98 \\ 
   8 & 0.05 &   2 & 7.26 & 8.23 & 8.29(3.05) & 9.37 & 0.05 & 0.06 & 0.07(1.47) & 0.08 & 2.27 & 2.85 & 3.36(4.77) & 3.87 \\ 
     & 0.10 &     & 6.32 & 7.26 & 7.40(2.80) & 8.36 & 0.08 & 0.10 & 0.12(2.39) & 0.13 & 2.25 & 2.84 & 3.17(4.53) & 3.72 \\ 
     & 0.25 &     & 5.07 & 5.97 & 6.16(3.30) & 7.13 & 0.17 & 0.22 & 0.32(5.51) & 0.29 & 1.97 & 2.62 & 2.83(5.81) & 3.43 \\ 
     & 0.05 &   5 & 6.72 & 7.64 & 7.76(2.96) & 8.83 & 0.05 & 0.06 & 0.08(2.56) & 0.08 & 3.35 & 4.43 & 5.05(5.17) & 6.11 \\ 
     & 0.10 &     & 5.73 & 6.74 & 6.91(3.05) & 7.99 & 0.08 & 0.10 & 0.13(3.16) & 0.13 & 3.29 & 4.29 & 4.82(5.58) & 5.81 \\ 
     & 0.25 &     & 4.79 & 5.61 & 5.81(3.39) & 6.69 & 0.15 & 0.19 & 0.29(5.84) & 0.26 & 3.01 & 4.17 & 4.58(5.88) & 5.59 \\ 
     & 0.05 &  10 & 6.11 & 7.06 & 7.14(2.96) & 8.14 & 0.05 & 0.06 & 0.08(2.32) & 0.08 & 5.37 & 7.50 & 8.30(7.99) & 10.22 \\ 
     & 0.10 &     & 5.15 & 6.19 & 6.33(3.35) & 7.24 & 0.08 & 0.10 & 0.12(2.68) & 0.12 & 4.93 & 7.04 & 7.88(8.25) & 9.84 \\ 
     & 0.25 &     & 4.19 & 5.05 & 5.23(3.71) & 6.19 & 0.14 & 0.19 & 0.28(5.45) & 0.26 & 4.57 & 6.51 & 7.60(9.81) & 9.54 \\ 
   \bottomrule
\end{tabular}}
\end{table}

\begin{table}[H]
\centering
\caption{\label{tab:sims2} Mean (m) and quartiles of the distributions of the local parameters $\hat{\boldsymbol{\psi}_i}= \{ \hat{\sigma}_i^2 ,\hat{\alpha}_i   ,\hat{\beta}_i \}$, averaged over the 200 simulated point patterns generated assuming an exponential form in both the spatial and temporal dimensions for the GFR with covariance as in \eqref{eq:nonsep}.}
\resizebox{\textwidth}{!}{\begin{tabular}{cccc|cccc|cccc|cccc|cccc}
  \toprule
      & True & & &  &  $\hat{\sigma}^2$  &  &  &  & $\hat{\alpha}$ & &     &  & $\hat{\beta}$  &     &  &  & $\hat{\delta}$  &     &   \\
\midrule
$\sigma^2$ & $\alpha$ & $\beta$ & $\delta$ & 25\% & 50\% & m (mse) & 75\% & 25\% & 50\% & m (mse) &  75\% & 25\% & 50\% & m (mse) &  75\% & 25\% & 50\% & m (mse) &  75\% \\ 
  \midrule
  5 & 0.05 &   2 & 1.80 & 4.99 & 7.25 & 9.60(6.69) & 13.94 & 0.03 & 0.05 & 0.12(0.51) & 0.07 & 0.16 & 0.38 & 0.71(3.33) & 0.96 & 1.01 & 1.84 & 1.49(1.07) & 2.00 \\ 
     & 0.10 &   5 & & 2.89 & 3.77 & 3.79(3.30) & 4.62 & 0.05 & 0.07 & 0.09(0.47) & 0.10 & 4.99 & 5.55 & 7.22(6.32) & 7.78 & 0.01 & 0.02 & 0.23(1.21) & 0.11 \\ 
     & 0.05 &   2 & 0.30 & 4.23 & 4.87 & 4.67(2.66) & 4.99 & 0.03 & 0.04 & 0.05(0.30) & 0.05 & 1.95 & 2.01 & 4.23(5.85) & 2.43 & 0.01 & 0.02 & 0.08(1.24) & 0.09 \\ 
     & 0.10 &   5 &  & 3.16 & 4.03 & 3.86(3.22) & 4.55 & 0.04 & 0.06 & 0.10(0.55) & 0.09 & 4.96 & 5.05 & 6.93(6.26) & 5.55 & 0.01 & 0.02 & 0.10(1.23) & 0.09 \\ 
    8 & 0.05 &   2 & 1.80 & 5.07 & 6.89 & 6.50(2.38) & 7.82 & 0.02 & 0.03 & 0.05(0.32) & 0.05 & 1.90 & 2.89 & 4.01(4.36) & 4.44 & 0.01 & 0.01 & 0.16(1.24) & 0.06 \\ 
     & 0.10 &   5 &  & 3.77 & 4.48 & 4.88(2.76) & 5.64 & 0.04 & 0.06 & 0.08(0.51) & 0.08 & 3.29 & 4.94 & 6.42(6.07) & 7.76 & 0.01 & 0.01 & 0.20(1.22) & 0.09 \\ 
     & 0.05 &   2 & 0.30 & 5.31 & 6.70 & 6.36(2.12) & 7.50 & 0.02 & 0.03 & 0.04(0.16) & 0.05 & 2.17 & 2.97 & 4.78(5.00) & 5.14 & 0.01 & 0.01 & 0.13(1.23) & 0.05 \\ 
     & 0.10 &   5 &  & 4.94 & 6.29 & 5.99(2.23) & 7.25 & 0.03 & 0.05 & 0.06(0.18) & 0.07 & 5.00 & 5.18 & 6.55(5.30) & 6.09 & 0.01 & 0.03 & 0.15(1.20) & 0.13 \\ 
   \bottomrule
\end{tabular}}
\end{table}

\section{Application to real seismic data}
\label{sec:application}

\begin{figure}[H]
    \centering
	\subfloat{\includegraphics[width=0.4\textwidth]{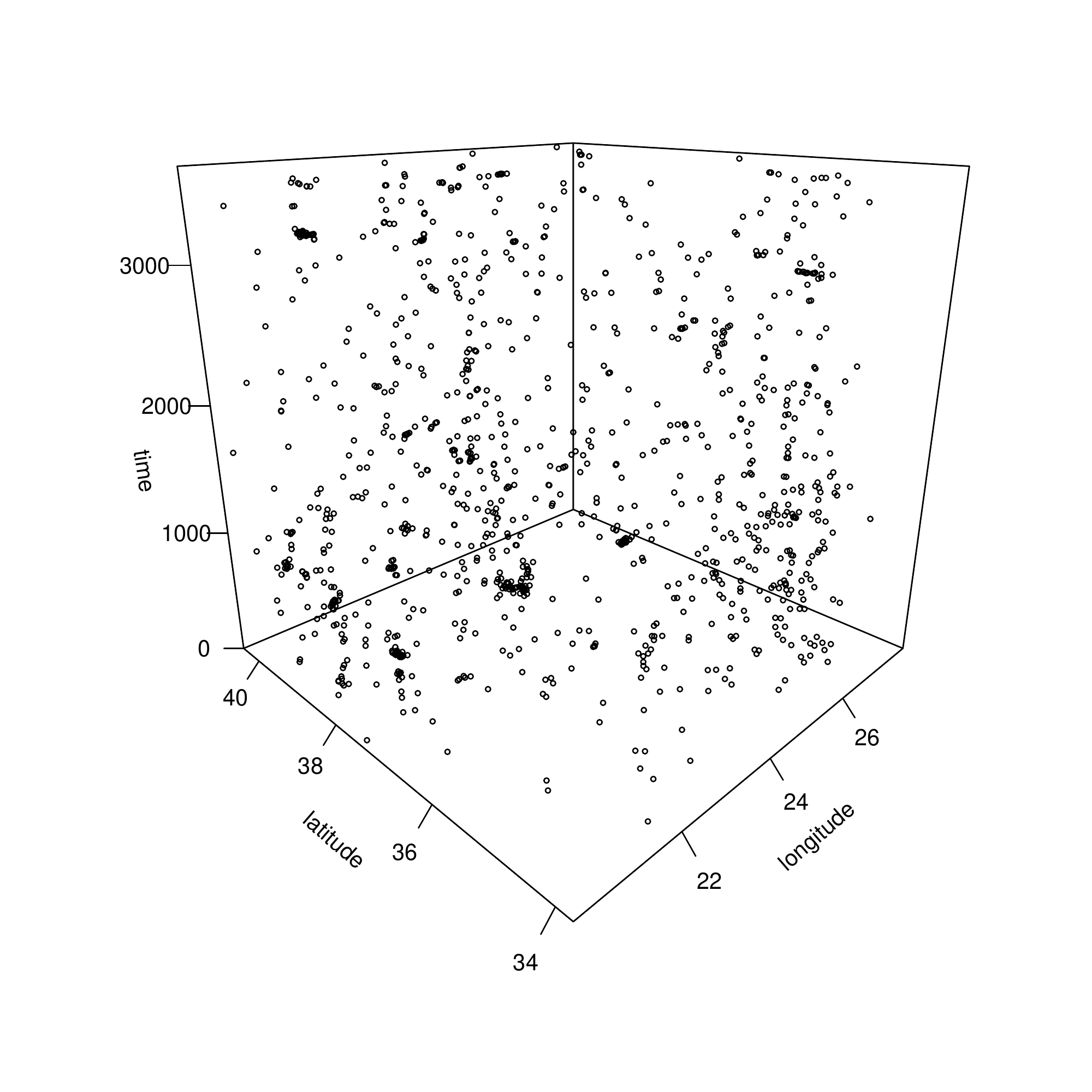}}  
	\subfloat{\includegraphics[width=0.4\textwidth]{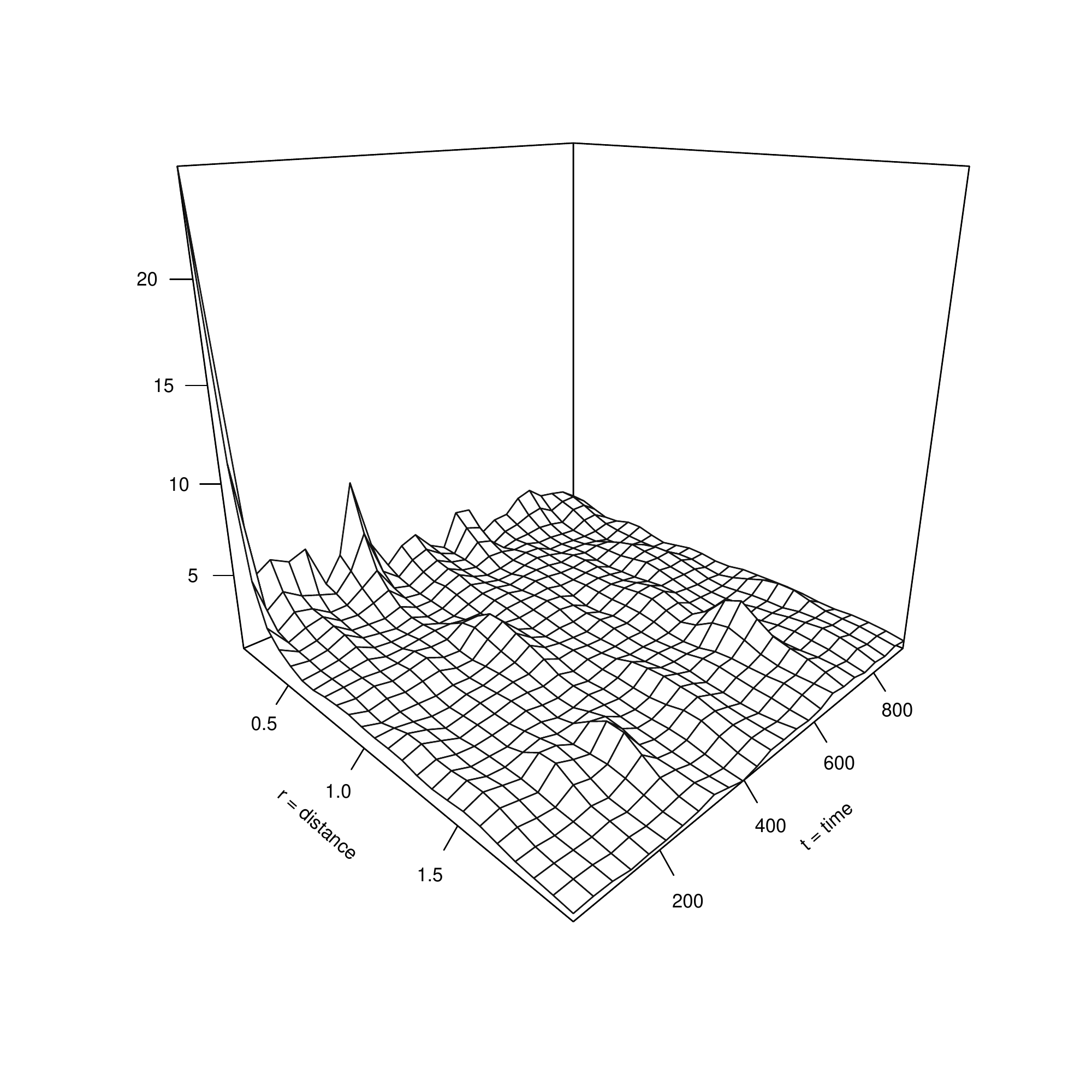}}
    \caption{Earthquakes occurred in Greece between 2005 and 2014 (a); Pair correlation function of the observed point pattern (b).}
    \label{fig:1}
\end{figure}

We analyse data related to 1111 earthquakes occurred in Greece  between 2005 and 2014, coming from the Hellenic Unified Seismic Network (H.U.S.N.). Time has been converted into days, and only seismic events with a magnitude larger than 4 are considered in this study.
The earthquakes are depicted in Figure \ref{fig:1}(a) , while Figure \ref{fig:1}(b) shows the pair correlation function \citep{stpp}.

The bandwidths $\epsilon$  and $\delta$, used in kernel estimation within the pair correlation function, are $0.15$ and $28.49$, for space and time, respectively.  Recall that the pair correlation function under complete randomness is a constant equal to one. 
Figure \ref{fig:1}(b) shows that the empirical pair correlation function takes values much larger than one for short distances indicating a tendency to clustering far from a Poisson process. Indeed we can also observe such clustered structures in Figure \ref{fig:1}(a).

We first consider a global LGCP with exponential covariance as in \eqref{eq:cov} using a joint estimation method as explained in Section \ref{sec:method1}. The corresponding spatial and temporal bandwidths are the same as indicated above in relation to Figure \ref{fig:1}.
We assume a constant intensity function estimated as $\hat{\lambda}(\textbf{u},t)=\hat{\lambda} =\frac{n}{|W \times T|} = 0.006 $.
The estimates of the covariance parameters are $\hat{\boldsymbol{\psi}}=\{  \hat{\sigma}^2,\hat{\alpha},\hat{\beta} \} = \{ 6.14 , 0.27 , 449.55\}$, which indicate a quite clustered underlying process, given the high estimated variance, and the relatively small scale parameters.

\begin{figure}[H]
    \centering
        \subfloat{\includegraphics[width=0.4\textwidth]{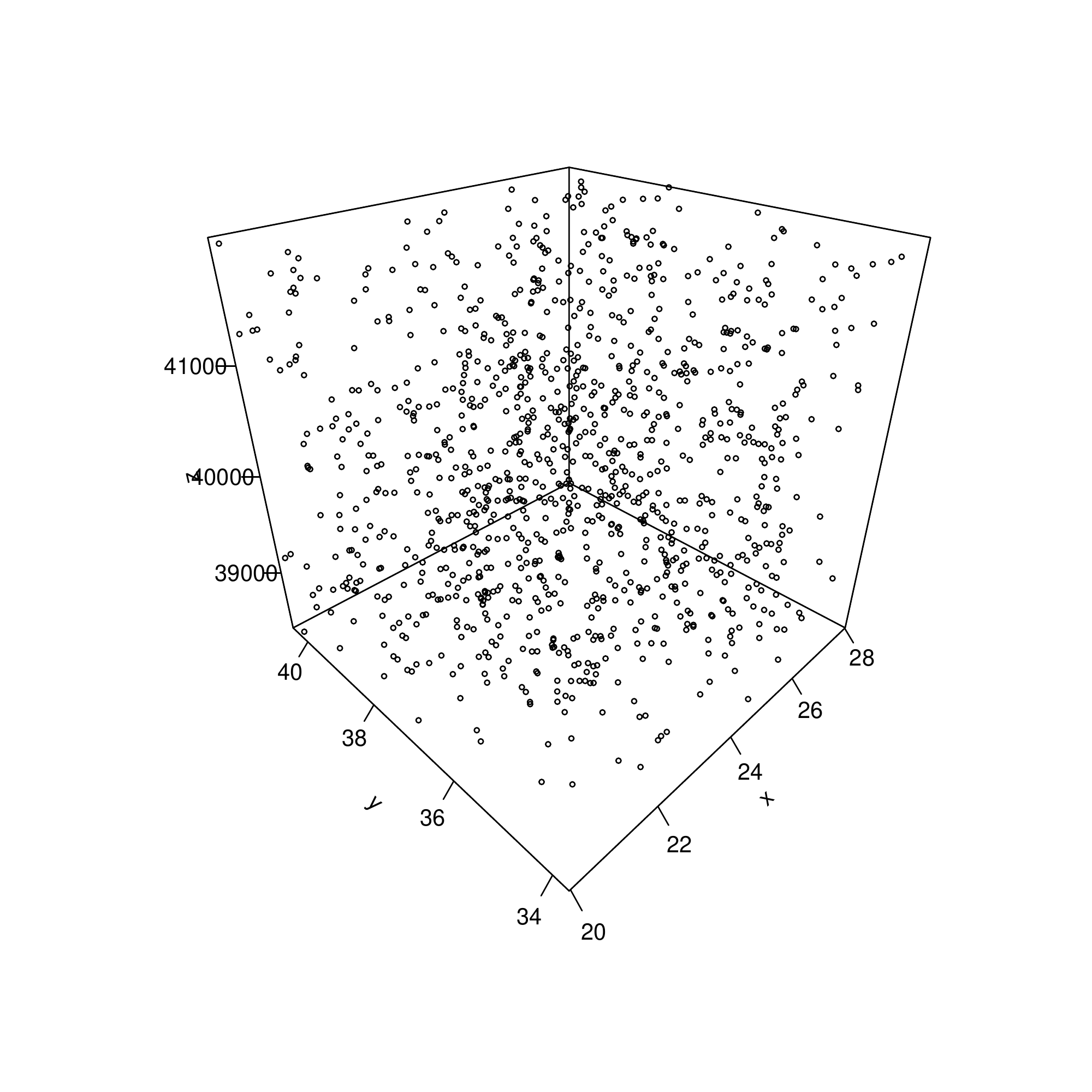}}
       \subfloat{\includegraphics[width=0.4\textwidth]{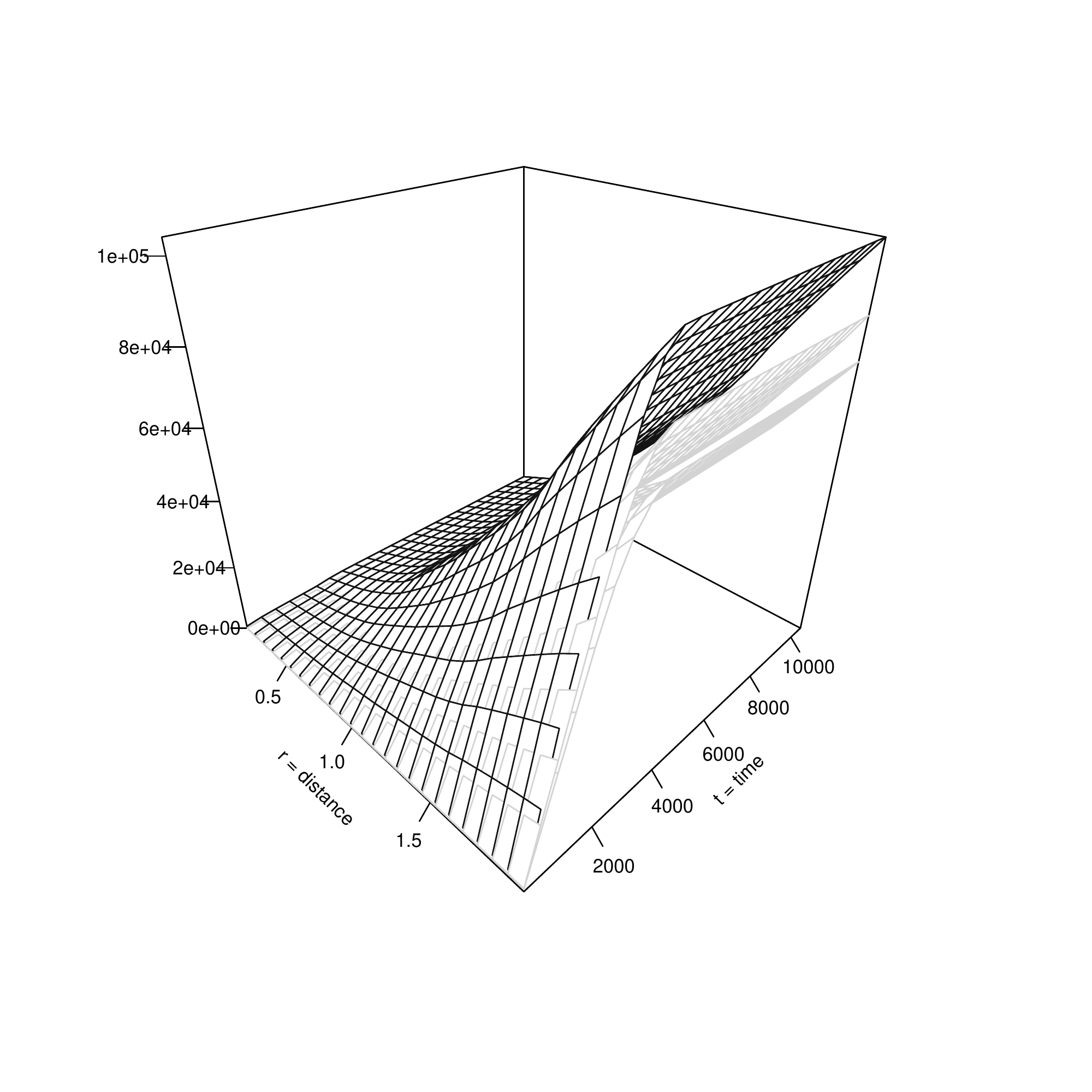}}
    \caption{(a) An example of LGCP point pattern simulated with the estimated parameters  $\hat{\boldsymbol{\psi}}$; (b) \textit{In black}: the estimated weighted $K$-function for the Greek seismic data. \textit{In light grey}: envelopes based on $39$ simulations from the global spatio-temporal LGCP at a significance level of 0.05.}
    \label{fig:kestglo}
\end{figure}

To assess the goodness-of-fit of the proposed (global) model, a residual analysis is carried out by means of the Monte Carlo test based on the inhomogeneous $K$-functions  outlined in Section \ref{sec:diagnostics}. Figure \ref{fig:kestglo}(a) displays an example of LGCP point pattern simulated with the estimated parameters  $\hat{\boldsymbol{\psi}}$, which is used to compute the envelopes for the test. Then, Figure \ref{fig:kestglo}(b) shows the the empirical $K$-function for the earthquake data, following \eqref{eq:kinh}, as well as the envelopes computed on $39$ simulations.
A overall p-value of 0 is obtained from \eqref{eq:pvalue}:
being equal to zero indicates that the empirical pattern is not compatible with the simulated ones (that come from a LGCP with estimated parameters $\hat{\boldsymbol{\psi}}$). Therefore,
the assumed model is not a good fit and an alternative should be searched, able to take into account the residual clustered behaviour of the points.
Furthermore, Figure \ref{fig:kestglo}(b) confirms this result, as the observed $K$-function does not lie within the envelopes.

\begin{figure}[H]
    \centering
	\subfloat{\includegraphics[width=0.4\textwidth]{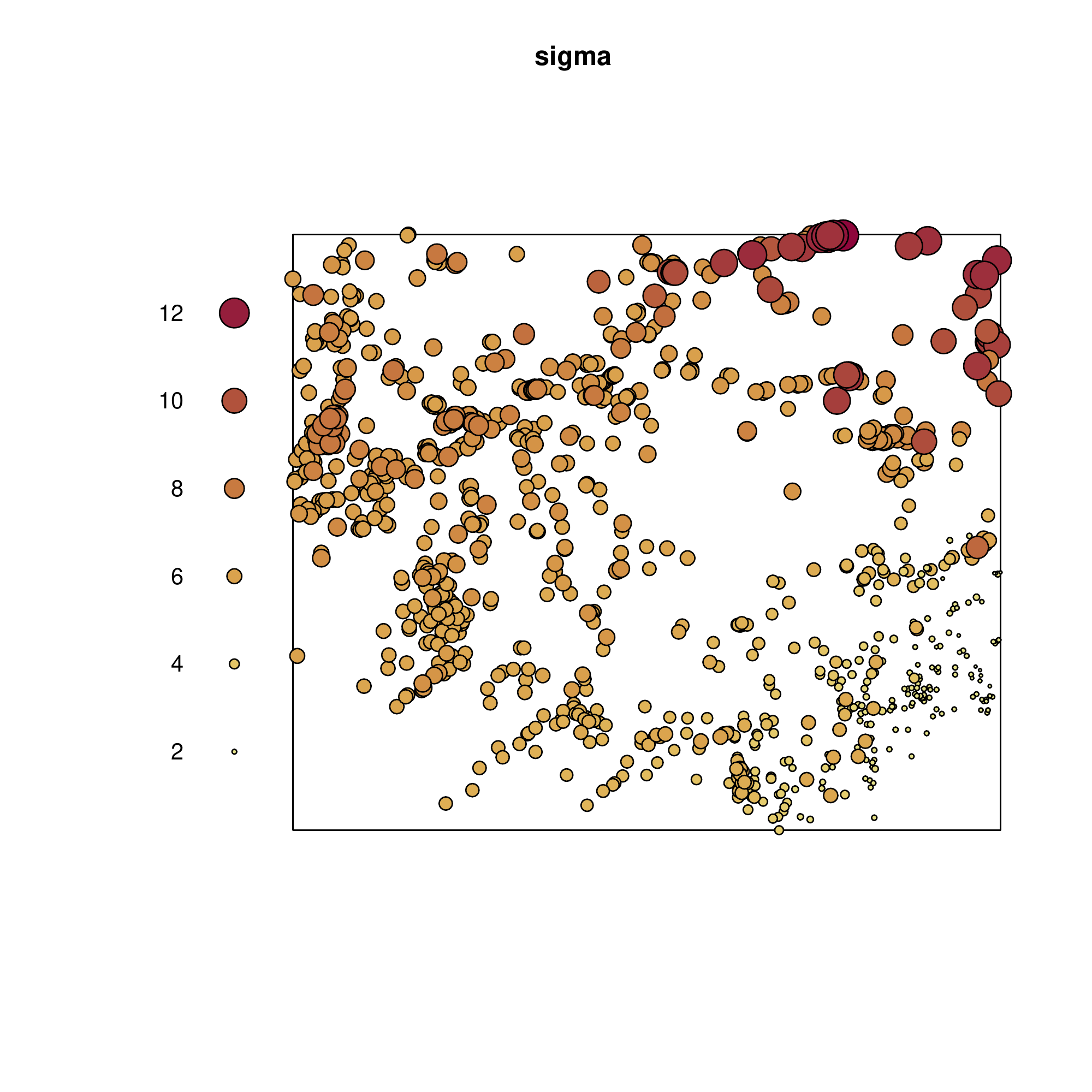}}
		\subfloat{\includegraphics[width=0.4\textwidth]{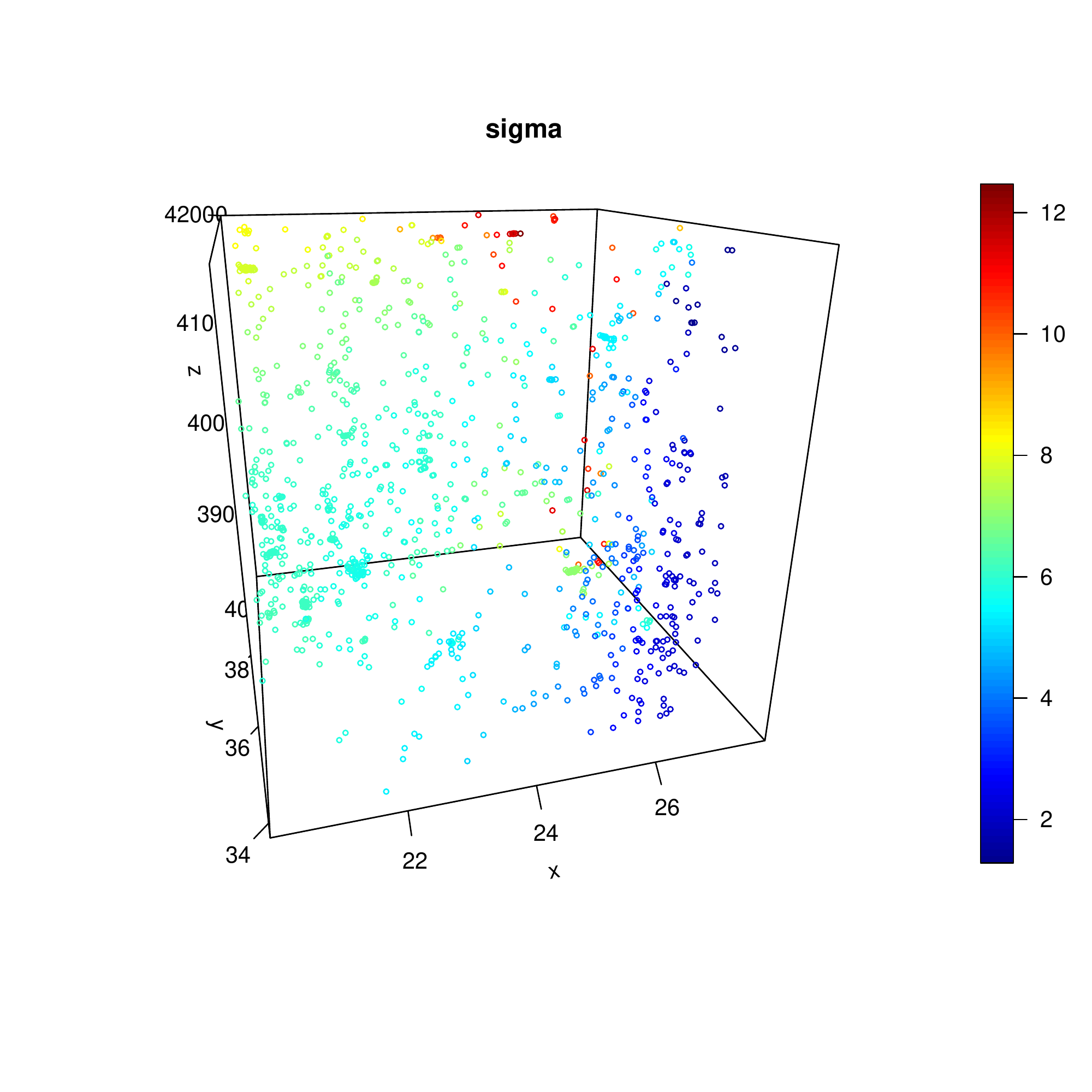}}\\
		\vspace{-.75cm}
		\subfloat{\includegraphics[width=0.4\textwidth]{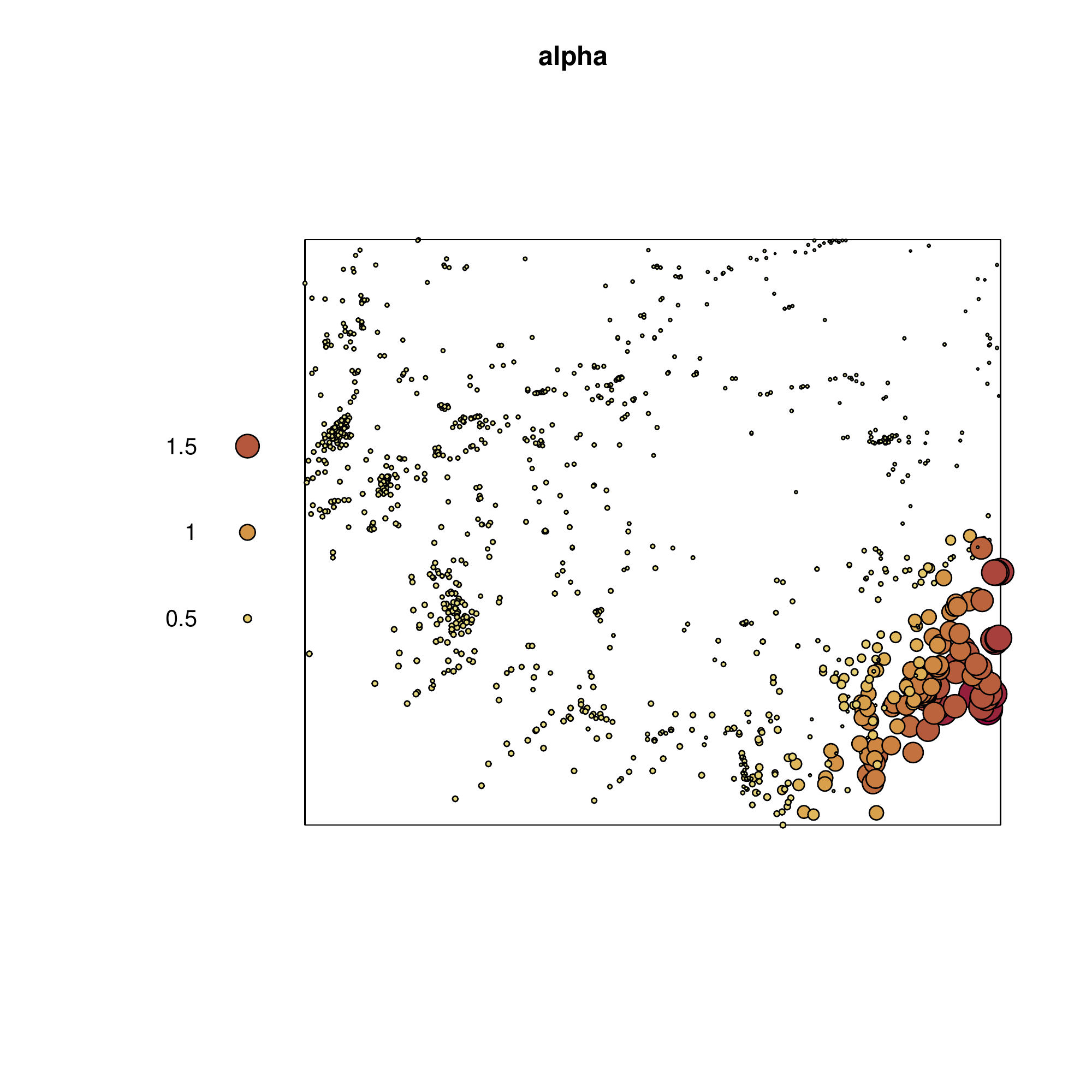}}
					\subfloat{\includegraphics[width=0.4\textwidth]{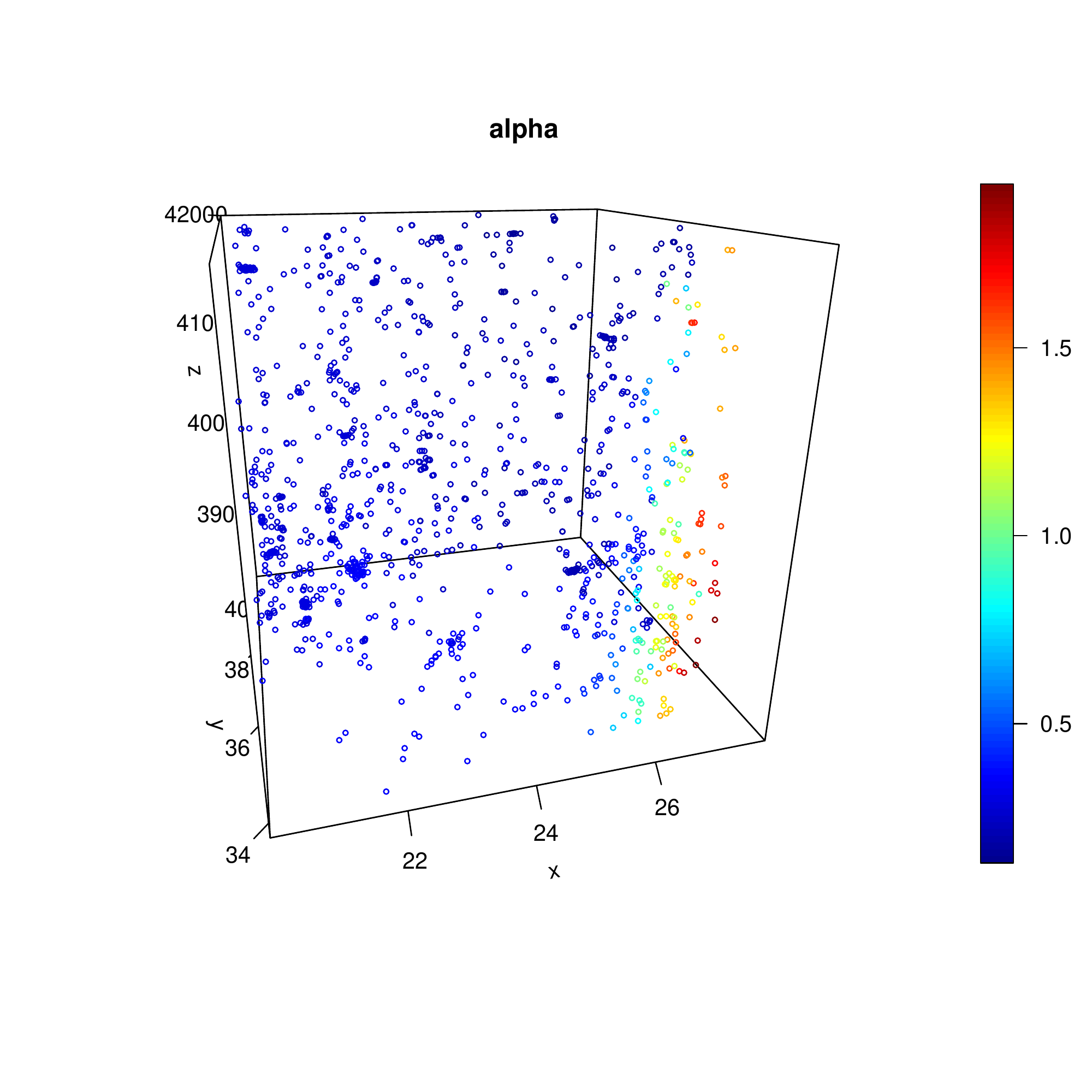}}\\
						\vspace{-.75cm}
			\subfloat{\includegraphics[width=0.4\textwidth]{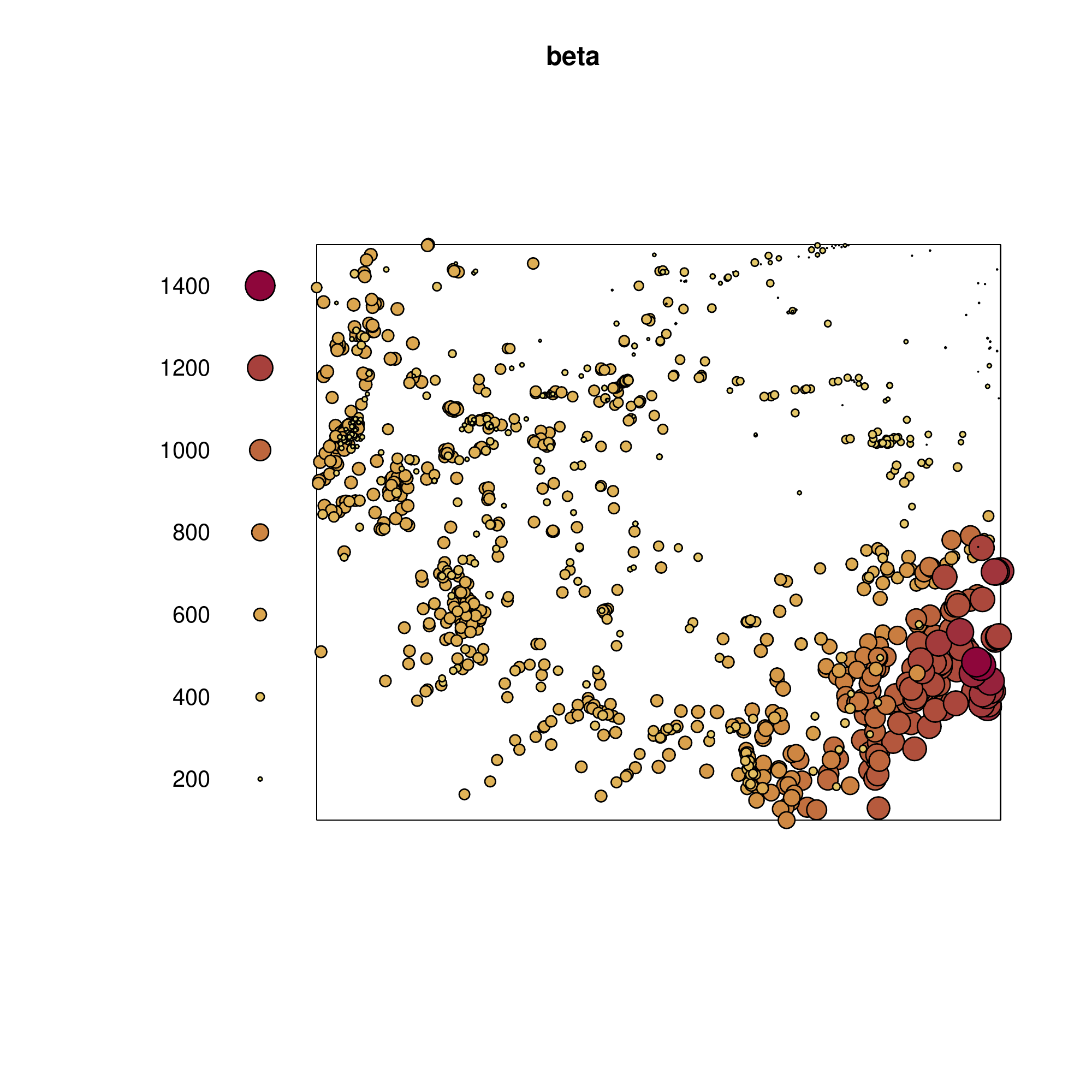}}
						\subfloat{\includegraphics[width=0.4\textwidth]{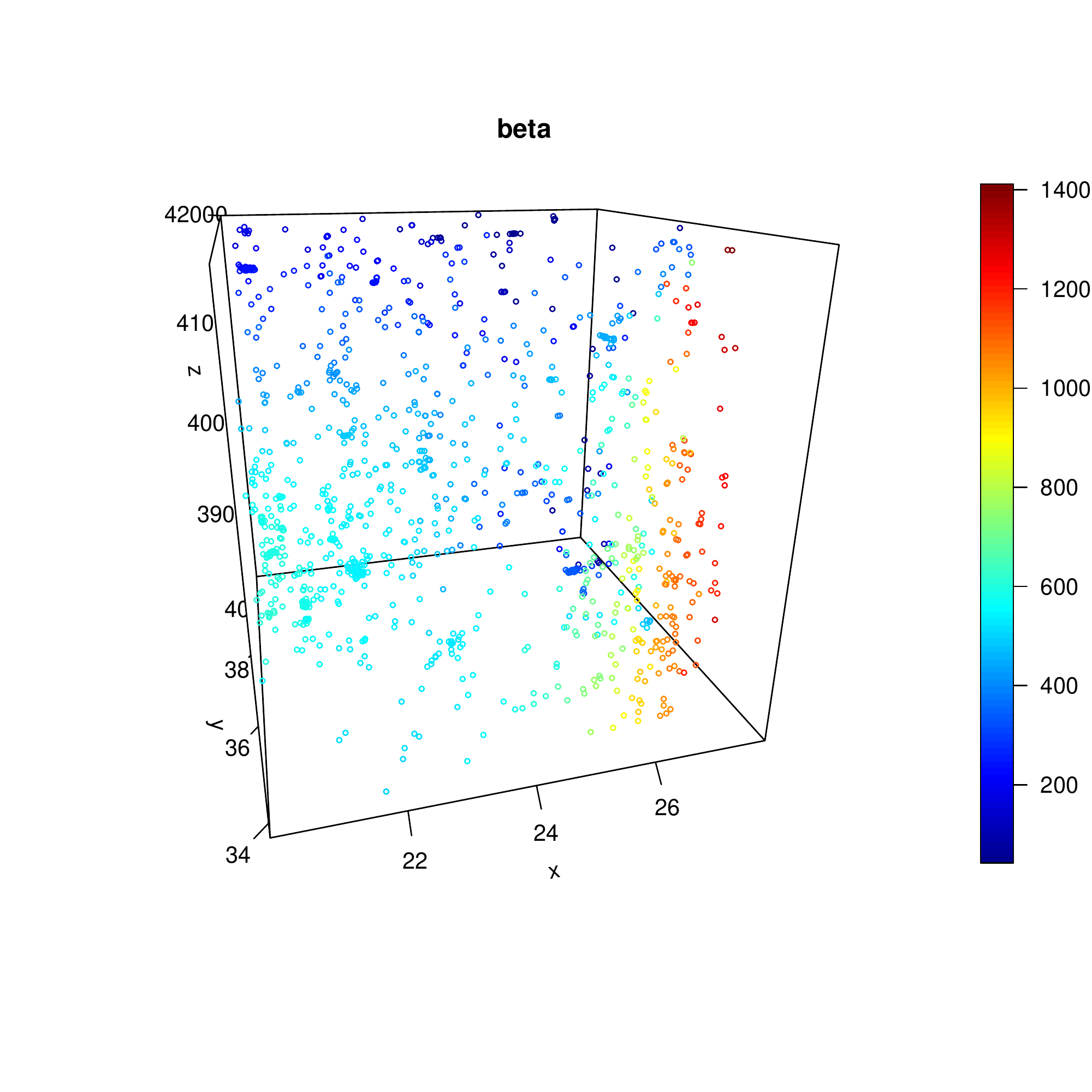}}
    \caption{Local estimates of the local LGCP fitted to the Greek seismic data. We show a spatial representation on the left and a spatio-temporal one on the right. 
    }
    \label{fig:3}
\end{figure}

We then consider a local spatio-temporal LGCP for the seismic data, and follow the  \textit{locally weighted minimum contrast} proposed in Section \ref{sec:proposall}.
The bandwidths $\epsilon$  and $\delta$, for the kernel in the pair correlation function, are the same used for the global fitting, while the bandwidths $\{ \sigma_x, \sigma_y, \sigma_t\}$ for the local weighting are estimated to be $2.58, 1.79$ and $1140.31$, for $x$, $y$, and $t$ coordinates, respectively.  
The results of the local fitting are shown in Table \ref{tab:1} and in Figure \ref{fig:3}.

\begin{table}[H]
\centering
\caption{Summary statistics of the estimated covariance parameters for the local LGCPs fitted to the Greek seismic data.}
\begin{tabular}{r|rrr}
  \toprule
 & $\hat{\sigma}^2$  & $\hat{\alpha}$ & $\hat{\beta}$ \\ 
  \midrule
Min. & 1.28 & 0.13 & 42.34 \\ 
  1st Qu. & 5.19 & 0.25 & 366.43 \\ 
  Median & 6.03  & 0.28 & 534.86 \\ 
  Mean & 5.77  & 0.39 & 535.52 \\ 
  3rd Qu. & 6.60 & 0.35 & 597.59  \\ 
  Max. & 12.47 & 1.94 & 1411.45 \\ 
   \bottomrule
\end{tabular}
\label{tab:1}
\end{table}

As shown in Table \ref{tab:1}, the proposed procedure provides a whole distribution for each parameter. Then, as evident from Figure \ref{fig:3}, the estimated parameters allow to clearly distinguish different sub-areas where points behave differently from each other. Indeed, by inspection of the left panels of Figure \ref{fig:3}, we can clearly spot two main spatial regions: the one on the top and on the left with higher $\hat{\sigma}_i^2$ values but lower estimates for $\hat{\alpha}_i$ and $\hat{\beta}_i$, and the area on the bottom-right which displays the opposite situation (low variance but large scale parameters).

This result is particularly appealing because it gives us more insight in the local behaviour of the seismic phenomenon. Indeed, from the global estimates we are only able to  draw the conclusion that the analysed process is overall clustered. Conversely, the local application shows that earthquakes occurred on the top-left of the analysed area are grouped in smaller clusters, while the ones on the bottom-left have a way more diffuse clustering behaviour. 

Furthermore, by looking at the right panels of Figure \ref{fig:3}, we can assess also the time-varying behaviour of the estimates and therefore on the underlying process: the most relevant result concerns the variability of the covariance parameters $\hat{\sigma}_i$ in time, indicating that the number of earthquakes tends to increase in time, and in particular, in the top-left spatial region.

\begin{figure}[H]
    \centering
	\subfloat{\includegraphics[width=\textwidth]{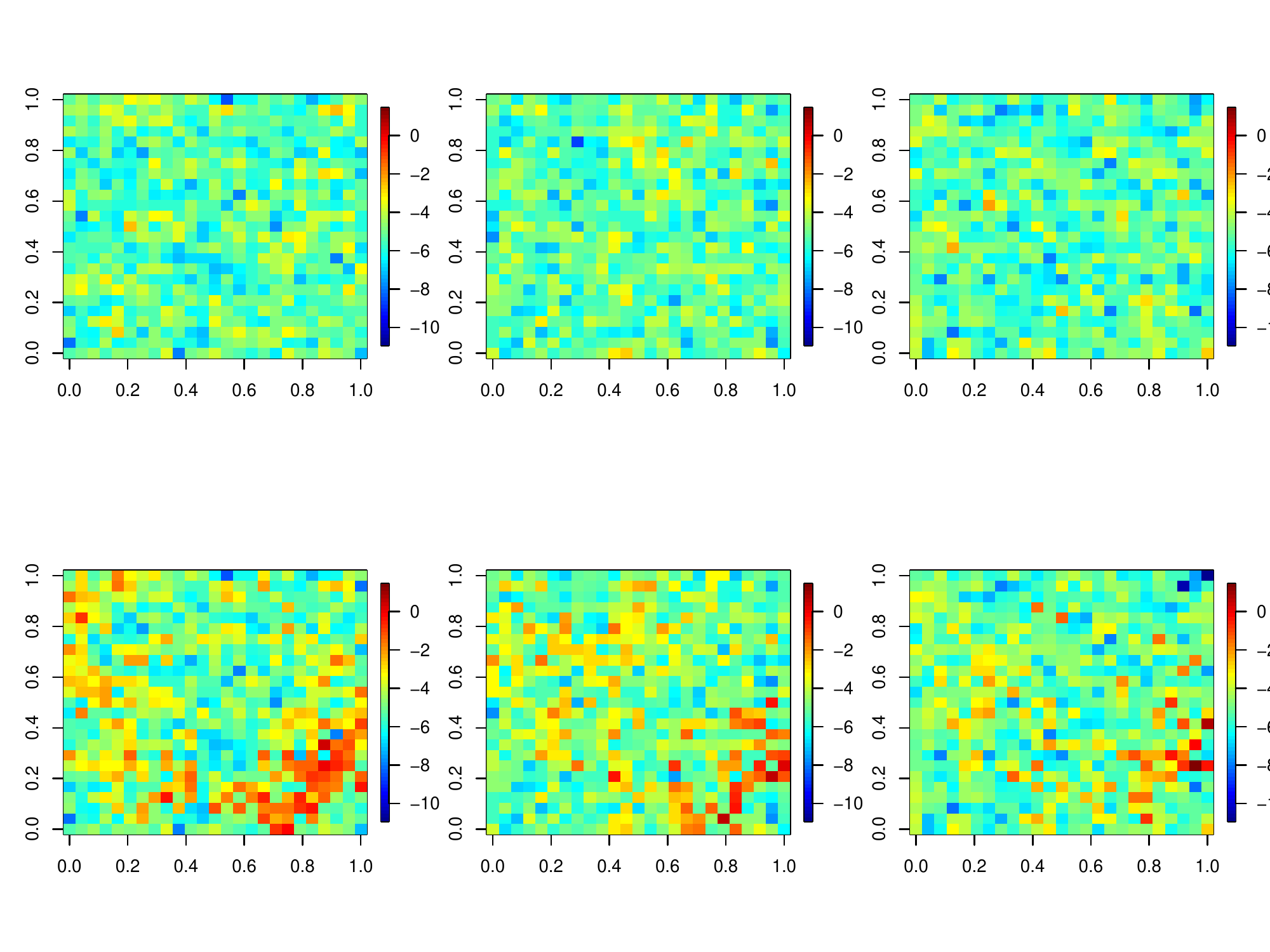}}\\
    \caption{GRFs for three temporal instants: \textit{Top panels:} global; \textit{Bottom panels:} local.}
    \label{fig:grfs}
\end{figure}

To assess whether the local model is a better fit to the data, also compared to the global counterpart, we proceed with the residual analysis as outlined in Section \ref{sec:diagnostics}. For the local LGCP, the procedure is based on the modification of the GRF to be used to generate the point patterns, allowing to include information of the local estimates.
To show the difference between the global and local GRFs, Figure \ref{fig:grfs} displays the spatial GRFs for three temporal instants, global and local, in the upper and lower panels, respectively. As evident, the local GRFs contain information about the local estimates, which in turn result in local variations.

\begin{figure}[H]
    \centering
        \subfloat{\includegraphics[width=0.4\textwidth]{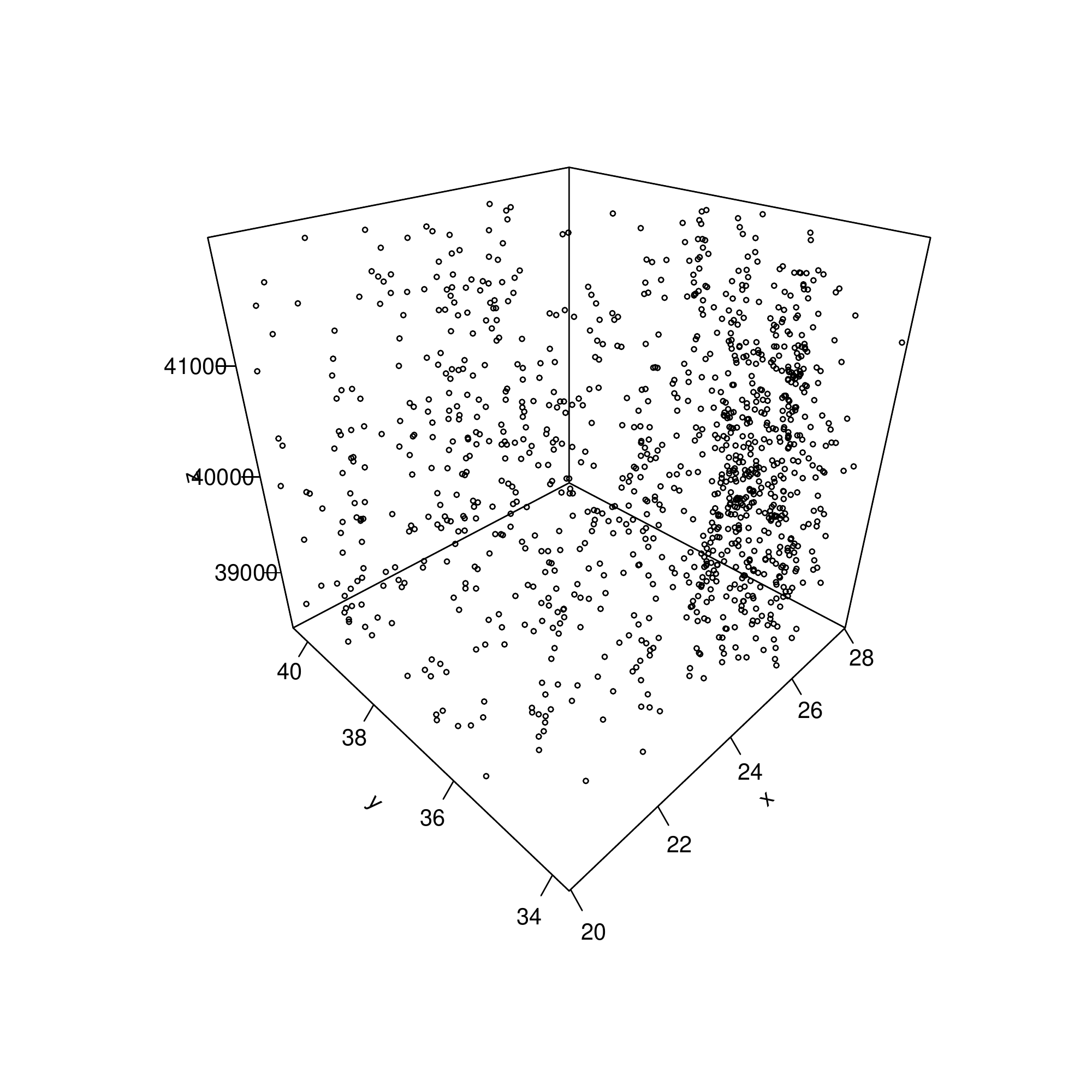}}
\subfloat{\includegraphics[width=0.4\textwidth]{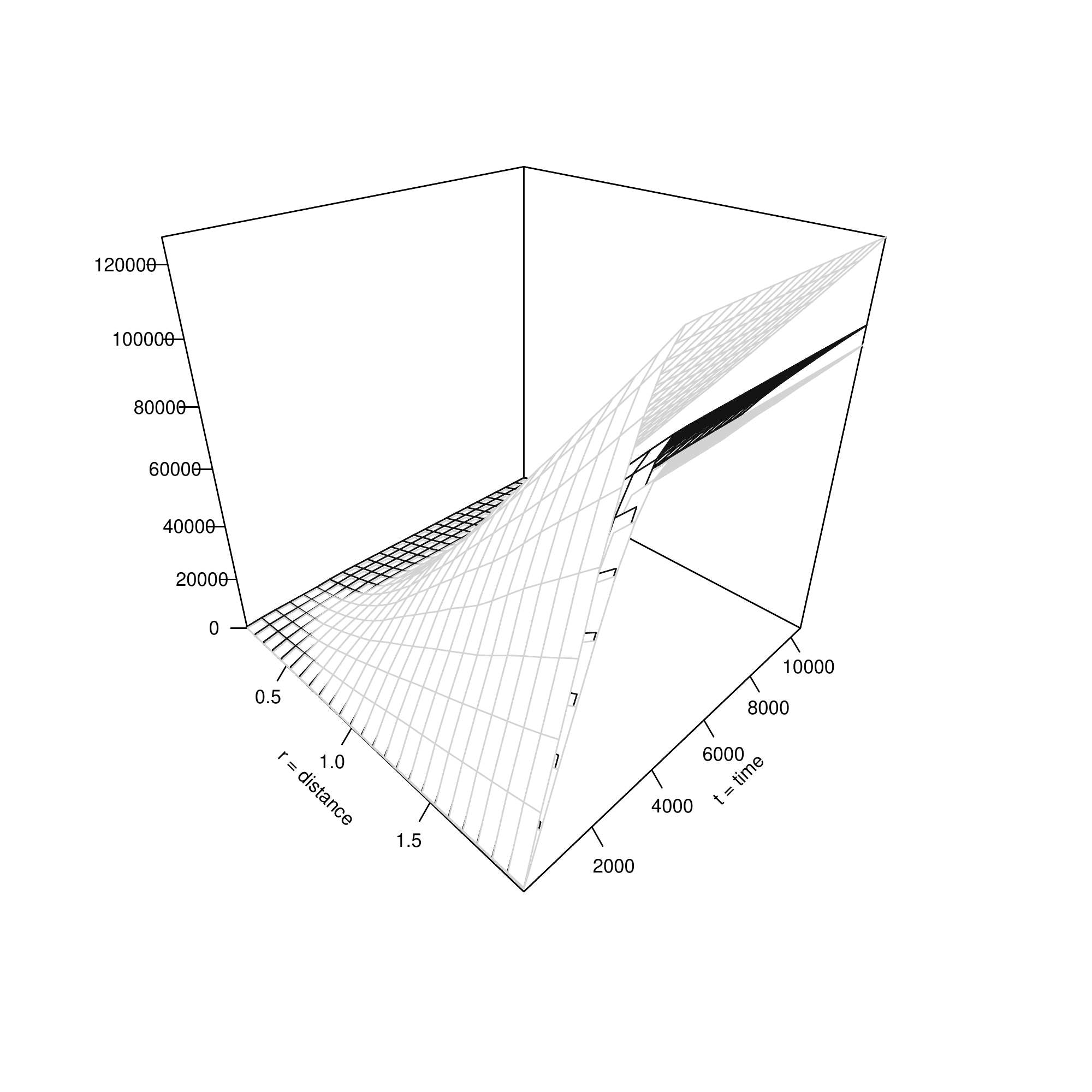}}
    \caption{(a) A point pattern simulated with the estimated parameters  $\hat{\boldsymbol{\psi}}_i$; (b) \textit{In black}: the estimated inhomogeneous $K$-function for the Greek seismic data. \textit{In light grey}: envelopes based on 39 simulations from the local spatio-temporal LGCP at a significance level of 0.05.}
    \label{fig:kestloc} 
\end{figure}

Finally, Figure \ref{fig:kestloc}(a) displays an example of point pattern simulated with the estimated local parameters  $\hat{\boldsymbol{\psi}}_i$, while  Figure \ref{fig:kestloc}(b) depicts the estimated $K$-function and the envelopes computed from $39$ local simulations. It is worth noticing how the point pattern simulated from the local parameters, following our procedure, better mimics the spatio-temporal arragement of the original points (Figure \ref{fig:1}(a)), if compared to the globally simulated point patterns (Figure \ref{fig:kestglo}(a)).
An overall p-value of $0.28$ is obtained using the Monte Carlo local test, indicating that the patterns simulated from the assumed model (with the local estimated parameters) do not present any residual clustering behaviour. Therefore, we can conclude that the local model is a good fit to the analysed data. This result is further confirmed by Figure \ref{fig:kestloc} (b), as now the observed $K$-function lies almost entirely between the envelopes obtained from the local simulation. Moreover,  the local model also represents a better fit than its previously fitted global counterpart.

\section{Conclusions}
\label{sec:conclusions}

We have introduced a novel local fitting procedure for obtaining space-time local parameters for a log-Gaussian Cox process.
From a methodological point of view, we have resorted to the joint minimum contrast procedure (which is appealing for its flexibility in dealing also with non-separable covariances), extending it to the local context, and therefore allowing to obtain a whole set of covariance parameters for each point of the analysed process.
The motivating problem came from the seismic application, where of course it is of interest studying the characteristics of the process in relation to both the spatial and the temporal occurrence of points.
By simulations, we have shown that the local proposal provides good estimates on average, if compared to the  global fitting alternatives.
Focussing on the application to real seismic data we have been able to assess that a local LGCP can be a better fit to the data if compared to its global counterpart. This is an expected result as the local fitting is based on a weighting given by a non-parametric estimate. However, our proposal presents further advantages than non-parametric alternatives, as it allows to obtain and interpret local parameters. 

Our proposal poses the basis for many further investigations and applications. Indeed, in the future, the proposed methodology could be applied to different real spatio-temporal point patterns, where it is of interest to study the characteristics of the underlying process, in relation to the spatial displacement and the temporal occurrence of points. Some examples may include seismology, forestry, criminology, or epidemiology.

Concerning the local LGCPs, it would be interesting to develop software to include external covariates into the first-order intensity and to propose diagnostic methods for the particular case of multiple covariates, as an extension to \cite{dangelo2021locall}.
Indeed, few spatio-temporal point process models account for external covariates: see \cite{adelfio2020including} for the ETAS model, and \cite{dangelo2021self} for the spatio-temporal Hawkes point process model adapted to events living on linear networks. However, to the best of our knowledge, none of the recent proposals provide local estimates of the model regression-type parameters.

The inclusion of covariates would further allow to explore and compare their effects both in global and local models. See for instance \cite{fotheringham2022importance}, which discuss  some examples of the spatial variant of the Simpson’s paradox, by local models for areal data.
Indeed, there might be some cases where the spatial variations observed in the local estimates of some covariates  could simply be due to some misspecification of the model form (such as omitting informative covariates), rather than to spatially inhomogeneous behaviour. The same issue could hold in some spatio-temporal contexts.

Moreover, future works could focus also on the use of other summary statistics, such as the $K$-function, into the local minimum contrast procedure. Suitable simulation studies could be carried out, and the performance of different local second-order summary statistics could be compared, as discussed in \cite{davies2013assessing}.

A related issue concerns the weights given to the averaged summary statistic, which strictly speaking could need not be kernel estimates nor Gaussian ones. For instance, \cite{zhuang2015weighted} (which applies the weighted likelihood estimator to the spatio-temporal ETAS model to study the spatial variations of seismicity characteristics in the Japan region) chooses a two-dimensional step-wise kernel function based on concentric disjoint octagon rings.
The optimal kernel function and bandwidth,  not discussed in detail in this study, could be investigated in future research.

Furthermore, other Cox models estimation could be carried out by exploiting the proposed minimum contrast procedure based on local second-order summary statistics.

\section*{Founding}
This work was supported by 'FFR 2021 GIADA ADELFIO'.

\bibliography{BBB}

\end{document}